\documentclass[twocolumn,prl,10pt,aps,floatfix]{revtex4-1}
\usepackage{graphicx}
\usepackage[colorlinks,linkcolor=blue,citecolor=blue,urlcolor=blue]{hyperref}
\usepackage{color}
\usepackage{subcaption}
\def\be{\begin{equation}}
\def\ee{\end{equation}}
\def\bea{\begin{eqnarray}}
\def\eea{\end{eqnarray}}
\def\nn{\nonumber}

\begin{document}

\title{Scattering of spinon excitations 
by potentials in the 1D Heisenberg model}
\author{A. Pavlis$^{1,2}$}
\author{X. Zotos$^{1,2,3}$}
\affiliation{$^1$ITCP and CCQCN, Department of Physics,
University of Crete, Herakleio, Greece}
\affiliation{$^2$Foundation for Research and Technology - Hellas, 71110
Heraklion, Greece}
\affiliation{
$^3$Leibniz Institute for Solid State and Materials Research IFW Dresden,
01171 Dresden, Germany}

\begin{abstract}
By a semi-analytical Bethe ansatz method and a T-matrix approach 
we study the scattering of a spinon, 
the elementary quantum many-body topological excitation 
in the 1D Heisenberg model, by local and phonon potentials. 
In particular, we contrast the scattering of a spinon
to that of a free spinless fermion in the XY model to highlight the 
effect of strong correlations.
For the one spinon scattering in an odd-site chain, 
we find a regular behavior of the scattering coefficients. 
In contrast, in an even-site chain there is 
a transfer of transmission probability between the two spinon branches 
that grows exponentially with system size.
We link the exponent of the exponential behavior to the dressed charge 
that characterizes the critical properties of the 1D Heisenberg model,
an interplay of topological and critical properties.
The aim of this study is a microscopic understanding 
of spinon scattering by impurities, barriers or phonons, 
modeled as prototype potentials, an input  
in the analysis of quantum spin transport experiments.

\end{abstract}

\maketitle

\section{Introduction}
The novel mode of thermal transport by magnetic excitations 
in quasi-one dimensional quantum magnets has been over the last few years the 
focus of extensive experimental \cite{hess} and 
theoretical studies \cite{xzth,dvsq,sk2002,rosch,louisk,sasha}.    
It was promoted by the fortuitous coincidence of synthesis of excellent quality 
compounds very well described by prototype integrable spin chain models 
and the proposal of unconventional -ballistic- spin and thermal transport 
in these systems \cite{dvsq}.
Of course the purely ballistic thermal transport predicted by theory 
is not observed in thermal conduction experiments as the, albeit very high, 
thermal conductivity is limited by the scattering of the magnetic 
excitations from impurities and phonons \cite{hess}.

In parallel, in the field of spintronics (spin caloritronics) 
there is renewed interest in the transport of magnetization, with 
the (inverse) spin Hall and spin Seebeck effect employed for the 
generation and detection of spin currents \cite{spintronics1,spintronics2}. 
So far mostly metallic, semiconducting and 
magnetically ordered (ferro, antiferro, ferri) magnetic materials 
have been studied. Only very recently the spin Seebeck effect 
was studied in the quasi-one dimensional 
quantum magnet Sr$_2$CuO$_3$ accurately described by a
spin-1/2 Heisenberg chain \cite{hirobe}.

Regarding quasi-one dimensional quantum magnets,  
a lot is known on their bulk thermodynamic \cite{ts}  and 
magnetothermal transport properties \cite{louis,sk,psaroudaki}.
The prototype model for these systems is the well studied 
1D Heisenberg model that is analytically solvable by the Bethe ansatz (BA) 
method. The elementary excitations in this strongly 
correlated system are topological in nature
- the spinons \cite{faddeev} -  
and most of thermodynamic and transport experiments are discussed 
in terms of these low energy excitations \cite{hess,rosch,sasha}.

In this work, we study the scattering of a spinon from  
local potentials aiming at a microscopic understanding 
of scattering processes by impurities, phonons and barriers, 
relevant to (far-out of equilibrium) quantum spin transport.
At the moment, we do not address any particular experiment, 
we only present background work on the 
theoretical question, {\it how does a 
quantum many-body topological excitation scatters from a potential ?}. 
This question is also 
relevant in other systems with topological excitations
of actual experimental and theoretical interest. 

To this end, we first use a recently developed semi-analytical 
Bethe ansatz method \cite{kitanine,caux} 
to evaluate scattering matrix elements by prototype potentials
and then to evaluate scattering coefficients by a T-matrix method.
We should emphasize that although it is an elementary exercise to 
evaluate the quantum mechanical scattering coefficients 
(reflection, transmission) of a free particle from a potential barrier,
little is known on the scattering of a quantum many-body 
quasi-particle excitation even more so for a topological one. 
The Bethe ansatz solvable models offer exactly such a framework for the study 
of this fundamental problem.

\section{Model and matrix elements}

The XXZ anisotropic Heisenberg Hamiltonian for a chain of $N$ sites
with periodic boundary conditions $S_{N+1}^a=S_1^a$ and in the presence 
of a local potential $V$ of strength $g$ is given by:

\begin{figure}[!ht]
\begin{center}
\includegraphics[width=1.0\linewidth]{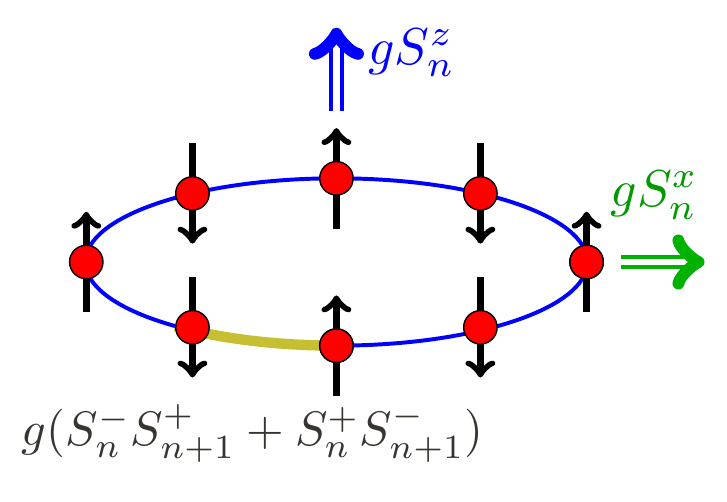}
\caption{Schematic figure of the spin chain and the type of potentials 
in consideration.} 
\label{fig1}
\end{center}
\end{figure}

\begin{eqnarray}
H&=&\sum_{n=1}^{N} h_{n,n+1} +gV
\\
h_{n,n+1}&=&J (S_n^x S_{n+1}^x+S_n^y S_{n+1}^y 
+\Delta S_n^z S_{n+1}^z-hS^z_n),
\nonumber
\label{heis}
\end{eqnarray}
where $S_n^a=\frac{1}{2}\sigma_n^a$,
$\sigma_n^a$ are Pauli spin operators with components
$a=x,y,z$ at site $n$,
h is the magnetic field and the anisotropy parameter $\Delta$ is typically 
parametrized as $\Delta=\cos \gamma$. In the following we will focus in the 
easy-plane antiferromagnet, $0 \le \Delta \le 1$ and we will take $J=1$ as
the unit of energy.

We study chains with odd as well as even number $N$ of spins.
In odd chains, for each total $S^z=\pm 1/2$,
the ground state is doubly degenerate containing one 
spinon with dispersion given by the one-branch 
$\varepsilon_Q=v_s\vert \sin Q\vert,~~~0 < Q < \pi$. 
For even N the lowest excitations involve at least two spinons,  
the dispersion of each spinon given by $\varepsilon_Q=v_s\vert \sin Q\vert$ 
i.e. states of the Cloizeaux-Pearson spinon spectrum 
\cite{cloizeaux, faddeev}. We will study states belonging to the 
lowest energy branch of the $M=N/2-1$ 
magnetization sector and obtained from the $S^z=1$ states 
by keeping the one spinon 
momentum fixed at zero and considering the dispersion of the second. 
In the spinon dispersion, $v_s=\frac{\pi}{2} \frac{\sin\gamma}{\gamma}$ 
and $Q$ is defined as the spinon momentum above the ground state. 
Normalized spinon states 
$\vert Q \rangle = \vert\lbrace\lambda\rbrace\rangle$ 
are determined from a specific 
set of Bethe roots $\lbrace\lambda_j\rbrace_{j=1}^M$, 
Supplemental Material \cite{suppl} (see, also, reference [1] therein), 
and matrix elements between such states describe spinon scattering processes. 
Moreover we define the spinon group velocity as $u_{Q}={d\varepsilon_{Q}/dQ}$. 

In the following we first evaluate scattering  matrix elements
$\vert{\cal M}\vert^2=\vert \langle Q'\vert V\vert Q\rangle\vert^2$ 
of a spinon from a state of momentum $Q$ to a state of momentum $Q'$ 
on finite size lattices following \cite{kitanine,caux,suppl}.
We show in particular that they are strongly enhanced 
compared to those of single particle excitations leading to 
unusual scattering coefficients. 
The potentials we consider are schematically shown in Fig.\ref{fig1}.

To start with we consider a one-site longitudinal potential 
$V=S_n^z$ at site $n$.
The corresponding matrix element is given by \cite{convention},
\begin{equation}
|\mathcal{M}^z_q(Q)|^2=|<Q+q|S^z_q|Q>|^2,
\label{msz}
\end{equation}
\noindent
where $S^z_n=\frac{1}{\sqrt{N}}\sum_q e^{-iqn} S^z_q$.
In the simple $\Delta =0$ case - XY model -
by a Jordan-Wigner transformation the spectrum corresponds to that of
free spinless fermions, $|M^z_q|^2=1/N$ and the potential moves only 
one fermion to a different state \cite{karbach}.   

\begin{figure}[!ht]
\begin{center}
\includegraphics[width=1.0\linewidth]{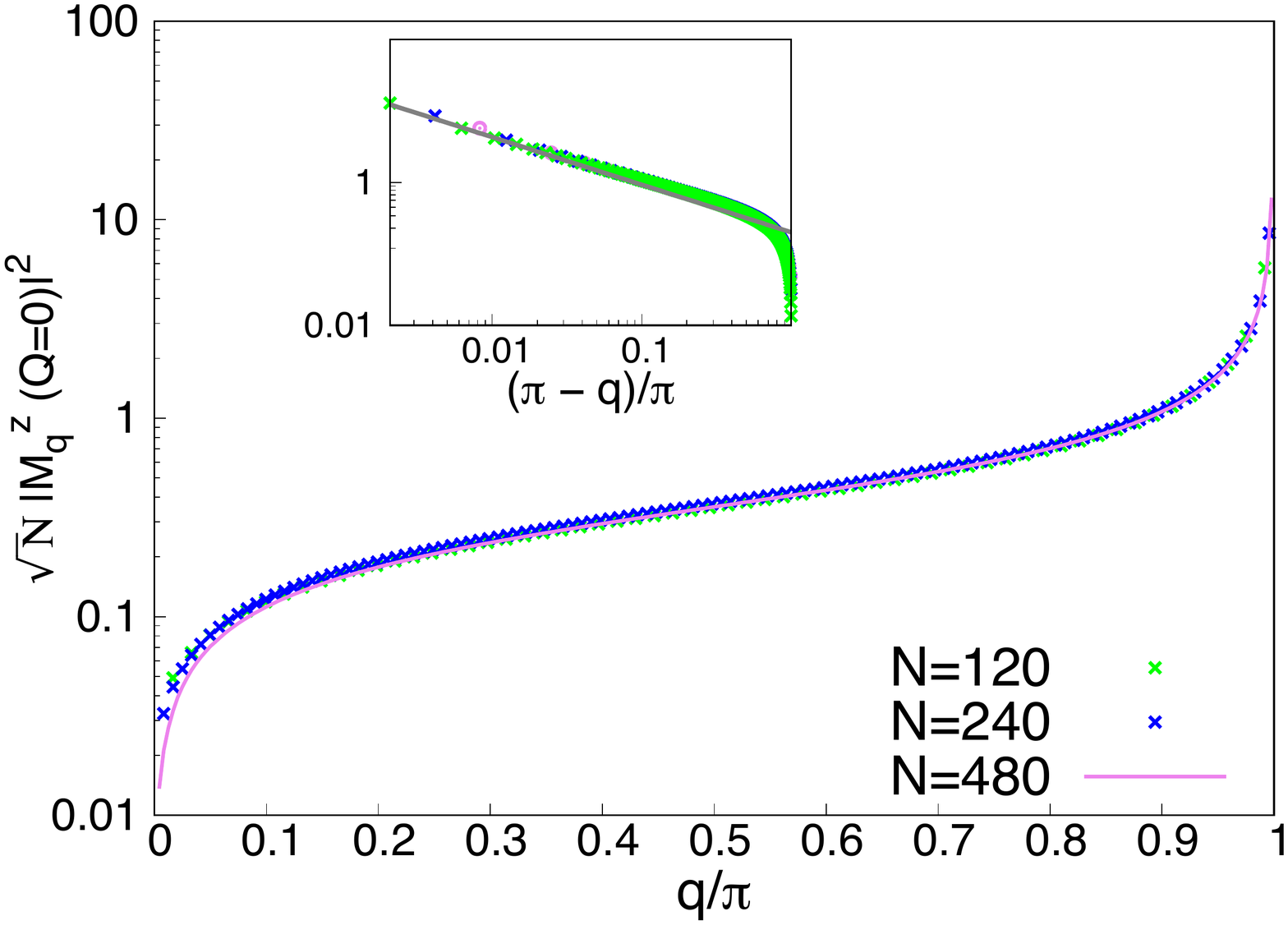}
\caption{Scaled $\sqrt{N}\vert\mathcal{M}^z_q(Q=0)\vert^2$ 
as a function of $q/\pi$ for $N=120,240,360,480$, $\Delta=1$.
In the inset the asymptotic scaling of $\vert\mathcal{M}^z_q(Q=0)\vert^2$ 
with a solid line indicating the asymptote $(\pi-q)^{2/3}$.} 
\label{fig2}
\end{center}
\end{figure}

In sharp contrast, in the isotropic Heisenberg model ($\Delta=1$), 
due to strong antiferromagnetic fluctuations, the scattering matrix elements
are drastically enhanced as shown in Fig.\ref{fig2}.
$\vert\mathcal{M}^z_q(Q=0)\vert^2$ scales in overall as $1/\sqrt N$ 
and as indicated in the inset of Fig.\ref{fig2} in the region not 
close to $q=0, \pi$ the matrix element behaves 
approximately as

\begin{equation}
\vert\mathcal{M}^z_q(Q=0)\vert^2 \sim \frac{1}{\sqrt N} 
\frac{1}{(\pi-q)^{2/3}}.
\label{mzq}
\end{equation}
Note that this behavior does not describe the $q=\pi$, which should 
not be diverging and scales differently with $N$, as will be discussed 
below.  
 
The most interesting part in Fig.\ref{fig2} and relation (\ref{mzq}) 
is that the matrix elements scale in a non-trivial fashion with $N$. 
In the XY model and for a $S^z_n$ potential all matrix elements scale 
as $1/N$ which is the usual case in lattice scattering. 
On the contrary for all $\Delta \ne 0 $ the matrix elements 
have a non trivial relation with respect to the spinon momentum and a 
particular scaling with respect to the number of spin sites, which is 
crucial to the spinon scattering. 

\begin{figure}[htp]
\begin{center}
\includegraphics[angle=0, width=1.0 \linewidth]{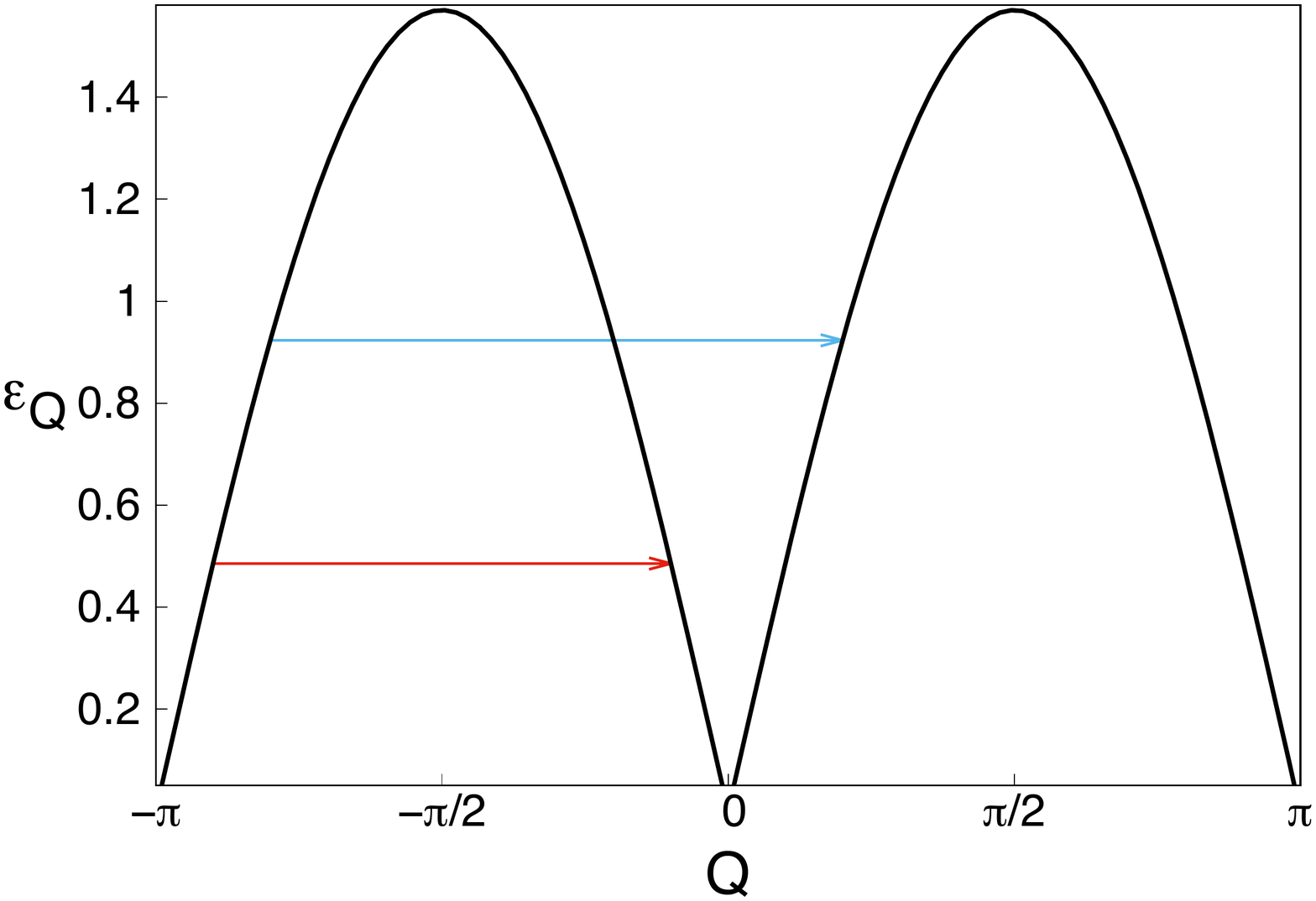}
\caption{Schematic description of the $Q\rightarrow\pi+Q$ transition (blue) 
and $Q\rightarrow \pi-Q$  same branch velocity flipping transition (red).}
\label{fig3}
\end{center}
\end{figure}

Furthermore, using \cite{kitanine} 
and a numerical evaluation, we further address the 
two types of matrix elements shown in Fig.\ref{fig3} 
(and all equivalent transitions between the two spinon branches)
that as we will see in the next section they play a significant role 
in the scattering processes. 
In the  $q=\pi$ transition 
\begin{equation}
\vert \langle Q+\pi\vert  
S^z_{\pi}\vert Q \rangle\vert^2\simeq \frac{f^z(Q)}{N^{2\mathcal{Z}^2-1} }
\end{equation}

\noindent
and in the same branch flipping velocity transition 
\begin{equation}
\vert \langle \pi-Q\vert  S^z_{\pi-2Q}\vert Q\rangle\vert^2\simeq
\frac{h^z(Q)}{ N^{\alpha(Q)}}, 
\end{equation}
\noindent
both corresponding to on-shell transitions. $\mathcal{Z}$ 
is the dressed charge introduced in \cite{korepin, bogoliubov} 
and the identification has been done using the analysis in 
\cite{gohman}, since  for small magnetic fields the dressed charge 
is $\mathcal{Z}\simeq \sqrt{{\pi\over 2(\pi-\gamma)}}$. 
In particular, $\mathcal{Z}^2=1$ for $\Delta=0$ and 
$\mathcal{Z}^2=1/2$ for $\Delta=1$. 
Note that this scaling of the matrix 
elements is also valid in the $h=0$ case, since by an analytical 
continuation the critical exponent $2\mathcal{Z}^2$ remains the same. 
Furthermore, for $Q$ not close to zero $\alpha(Q)\simeq 1$ and
$f^z(Q)$  is an almost constant function, while $h^z(Q)$ 
is a rapidly decreasing one to a constant value \cite{suppl}. 
These types of matrix elements have 
been extensively studied in \cite{kitanine1,kitanine2} and the correspondence 
between the dressed charge and the scaling of the matrix elements 
has been proven analytically.

\begin{figure}[htp]
\begin{center}
\includegraphics[angle=0, width=1.0\linewidth]{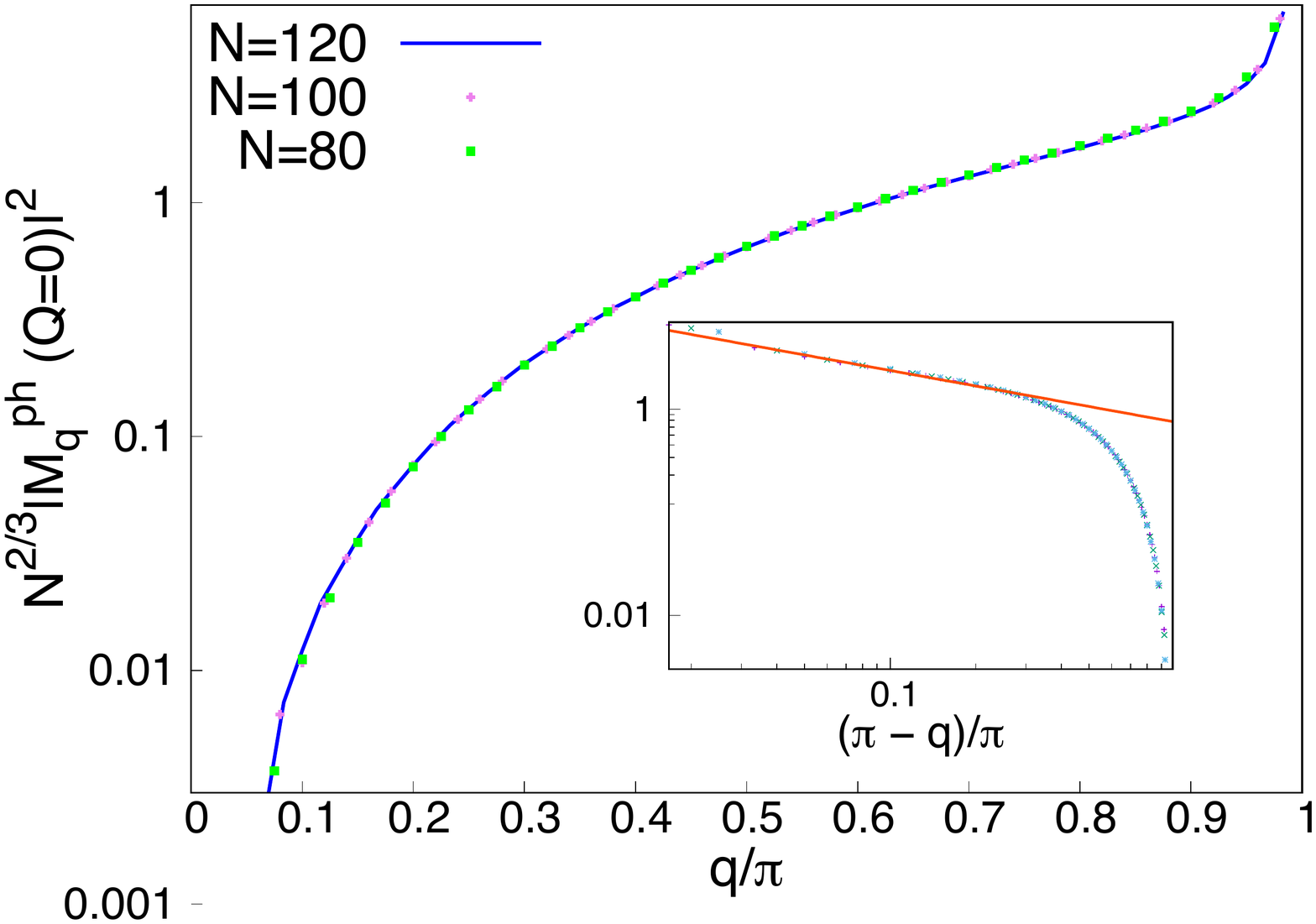}
\caption{Scaled $ N^{{2\over 3}} \vert 
\mathcal{M}^{ph}_q(Q=0)\vert^2$ as a function of $q/\pi$. 
The inset shows the asymptotic scaling of 
$\vert\mathcal{M}^{ph}_q(Q=0)\vert^2$ as a function of $(\pi-q)/\pi$. 
The solid line indicates the asymptote $(\pi-q)^{1/2}$. } 
\label{fig4}
\end{center}
\end{figure}

Next we consider the scattering of a spinon by a lattice distortion of 
wave-vector $q$,
\begin{eqnarray}
h_q&=&{1\over \sqrt{N}}\sum_{n=1}^{N} e^{iqn} 
J (S_n^x S_{n+1}^x+S_n^y S_{n+1}^y )
\label{hq}
\end{eqnarray}

\noindent
from which we can deduce the scattering from a "weak link" 
$V=g(S^{-}_nS^{+}_{n+1}+S^{+}_nS^{-}_{n+1})$. 
Similarly to the previous case, the scaled scattering matrix element 
for $\Delta=1$ and the asymptotic form 
\begin{equation}
|\mathcal{M}^{ph}_q(Q=0)|^2\sim  N^{-2/3}\frac{1}{(\pi-q)^{1/2}},
\end{equation}
\noindent
for $Q=0$ are shown in Fig.\ref{fig4}. 

Again, the dominant matrix elements for spinon 
scattering are 
a $\pi$-transition and a same branch velocity flipping matrix element,
\bea
\vert \langle Q+\pi\vert  h_{\pi}\vert Q\rangle\vert^2&\simeq&  
{f^{ph}(Q)\over N^{a(Q)} },
\nonumber\\
\vert\langle  \pi-Q\vert  h_{\pi-2Q}\vert Q\rangle\vert^2&\simeq&
{h^{ph}(Q)\over N}. 
\label{domph}
\eea
For the isotropic $\Delta=1$ case, $a(Q)$ has a weak 
dependence with respect to $Q$, $a(Q)\simeq 0.4$ 
around $Q=2\pi/10$, while by a Jordan-Wigner 
transformation we can derive that for $\Delta=0$ the 
absolute value squared of all matrix elements scales as $1/N$.

Finally we consider a transverse magnetic potential, $V=g S^x_n$. 
The main difference of this 
potential to the two previous ones is that 
it acts non-trivially only between states with $\Delta S^z=\pm 1$. 

\begin{figure}[h!]
\centering
\includegraphics[width=1.0\linewidth]{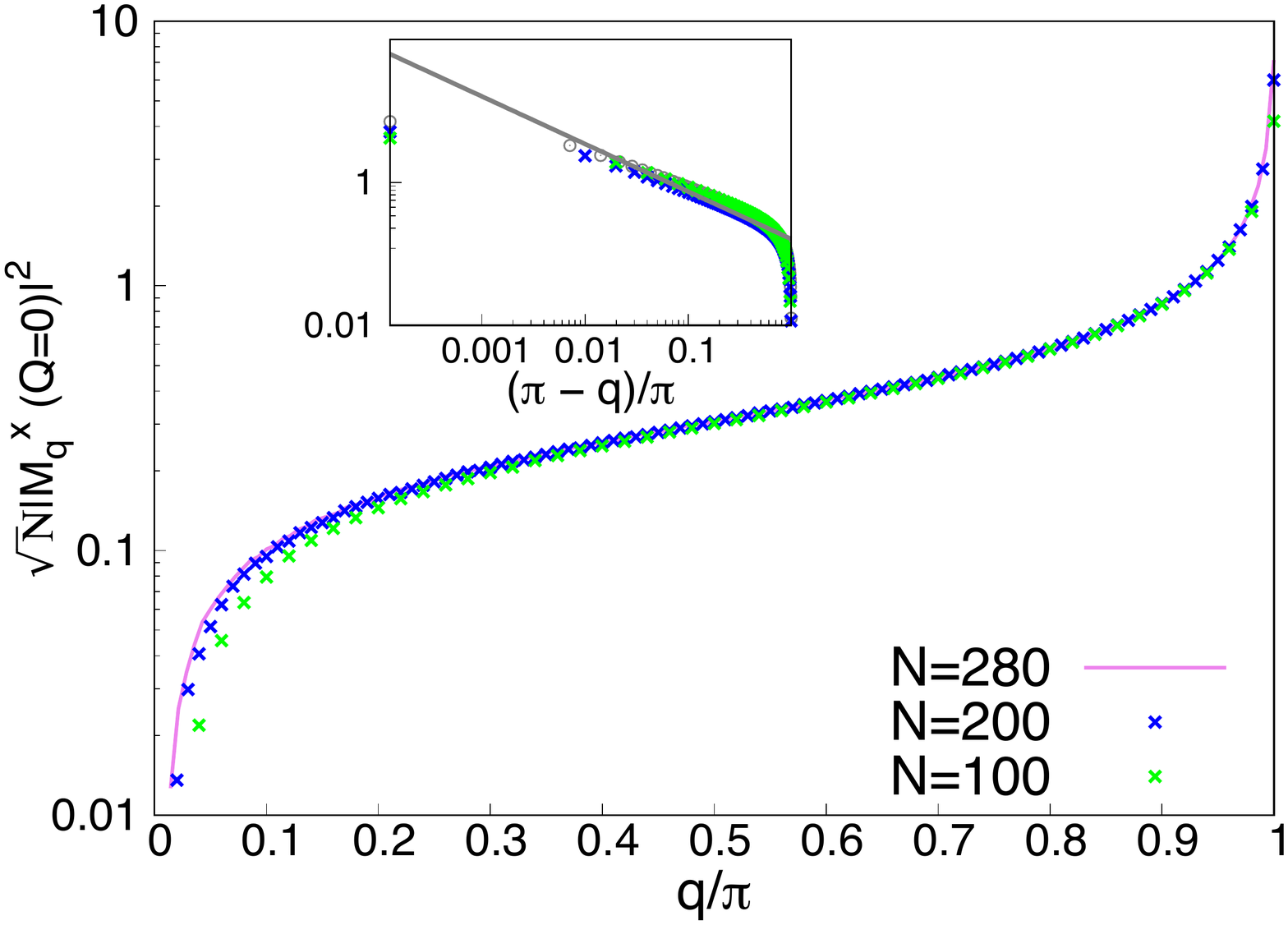}
\caption{Scaled $\vert M^x_q(Q=0)\vert^2 \sqrt{N}$ vs $q$ 
for the isotropic model $\Delta=1$ 
and various $N$. The solid line in the inset shows that the asymptote 
scales as $(\pi-q)^{2/3}$. }
\label{fig5}
\end{figure}

Similarly to the $S^z_n$ potential, as shown in Fig.\ref{fig5}, 
the asymptote behaves as,
\be 
\vert M^x_q(Q=0)\vert^2 \sim {1\over\sqrt{N}}{1\over (\pi-q)^{2/3}}
\ee

\noindent
and the dominant matrix elements scale as,

\be 
\vert \langle Q+\pi\vert  S^x_{\pi}\vert Q\rangle\vert^2\simeq 
{f^x(Q)\over N^{{1\over 2\mathcal{Z}^2}-1} }.
\ee

\noindent
This time, the XY model matrix elements behave 
non-trivially as they scale as $\sqrt N$ 
and in fact they imply a strongest scattering compared 
to the $0 < \Delta\leq 1$ case. 

Overall, the $\pi$-transitions show a strong $N-$dependence and a weak
$Q$-dependence, while the $\pi-2Q$ (velocity flip)
transitions show a $1/N$-dependence and a strong $Q-$dependence \cite{suppl}.

To close our discussion on the matrix elements, we consider
an extended potential profile $V_{ext}=\sum_{n=1}^N g_n V_n$, where $V_n$ 
represents one of the potentials we studied above and $g_n$ is the 
potential profile
\be 
\vert \langle Q+q\vert V_{ext}\vert Q \rangle\vert^2 
= {1\over N}\vert\sum_{n=1}^N g_n e^{-i q n} \vert^2 \vert V_q\vert^2.
\ee
For example, for a segment of $m$-sites with a
potential $V_m=\sum_{n=N/2}^{N/2+m-1} S_n^z$ the matrix element is given by,
\begin{equation}
|<Q+q|V_m|Q>|^2=
\frac{1}{N} \frac{ \sin^2\frac{qm}{2} } { \sin^2 \frac{q}{2} }
|\mathcal{M}^z_q(Q)|^2.
\label{vm}
\end{equation}
\noindent
This form of equation can be interpreted as a "diffraction"-like
pattern modified by the scattering of the spinon. For the XY model 
it simply becomes,

\begin{equation}
|<Q+q|V_m|Q>|^2= \frac{1}{N^2}
\frac{ \sin^2\frac{qm}{2} } { \sin^2 \frac{q}{2} }.
\end{equation}

The main message of this section is that the scattering matrix 
elements of the quantum many-body topological - spinon -  
excitations in the XXZ Heisenberg model 
are strongly enhanced compared to the ones in the XY model (free fermions).
They show a nontrivial system size dependence and thus we expect profound 
differences in the scattering of spinon excitations by a potential to the 
generic single particle one.

\section{Scattering coefficients}

We will analyze the transmission/reflection 
scattering coefficients of a spinon from a potential 
within the T-matrix approach by writing all quantities in the basis 
of Bethe ansatz eigenstates $\vert\lbrace\lambda\rbrace\rangle$, 
\bea
T&=&V\frac{1}{1-G_0 V}
\nonumber\\
G_0(E)&=&\lim_{\varepsilon\rightarrow 0}\sum_{\lbrace\lambda\rbrace} 
{\vert\lbrace\lambda\rbrace\rangle\langle \lbrace\lambda\rbrace\vert 
\over {E-E_{\lbrace\lambda\rbrace}+i\varepsilon}}\nonumber\\
V&=&\sum_{\lbrace\mu\rbrace,\lbrace\lambda\rbrace} 
\langle\lbrace \lambda\rbrace\vert V\vert \lbrace \mu\rbrace\rangle 
\vert \lbrace\lambda\rbrace\rangle\langle\lbrace\mu\rbrace\vert.
\eea

\noindent
$E_{\lbrace\lambda\rbrace}$ is the 
energy corresponding to the Bethe state $\vert \lbrace\lambda\rbrace\rangle$. 
Based on the discussion in the previous section 
for the particular scaling of the matrix elements with $N$, 
we write a typical matrix element in the form 
$\langle\lbrace\lambda\rbrace\vert V\vert \lbrace\mu\rbrace\rangle
=g f_{\lbrace\lambda\rbrace,\lbrace\mu\rbrace}/N^{\alpha}$  with $g$ 
being the potential strength and $\alpha=\alpha(\lbrace\lambda\rbrace,
\lbrace\mu\rbrace)>0$ a scaling factor. The potential matrix $V$ 
belongs in a Hilbert space of dimension  
$\dim\mathcal{H}=2^N$ which makes the problem intractable from a 
computational point of view. Therefore, in order to be able to calculate 
the scattering coefficients for relatively long spin chains, we restrict 
our numerical calculations to including only the two-spinon continuum 
i.e. a subspace of dimension $\dim\mathcal{H}_{2sp} = {N\over 8}(N+2)$. 
The calculation of the T-matrix is straightforward, 
we compute the matrix $1-G_0V$ and subsequently invert it and left multiply 
it by $V$. Note that for the evaluation of the Green's function $G_0$ 
we use the identity 
$\lim_{\varepsilon\rightarrow 0}{1\over{x+i \varepsilon}}
=P{1\over x}-i\pi\delta(x)$, 
where $P$ stands for the Cauchy principal value part.

\subsection{"Free" spinon}

It is instructive to consider the scattering of a free 
particle on a lattice with a "spinon" dispersion relation 
$\varepsilon_Q=v_s\vert \sin Q\vert$ 
by a one-site $\delta$-like potential of strength $g$. 
In this case all the matrix elements are the same, 
$<Q'|V|Q>=g/N$ and the transmission coefficient $\mathcal{T}_{Q,Q}$ 
is a function of $g/u_Q$ \cite{suppl}, $u_Q=d\varepsilon_Q/dQ$.

\begin{figure}
\centering
\includegraphics[width=1.0\linewidth]{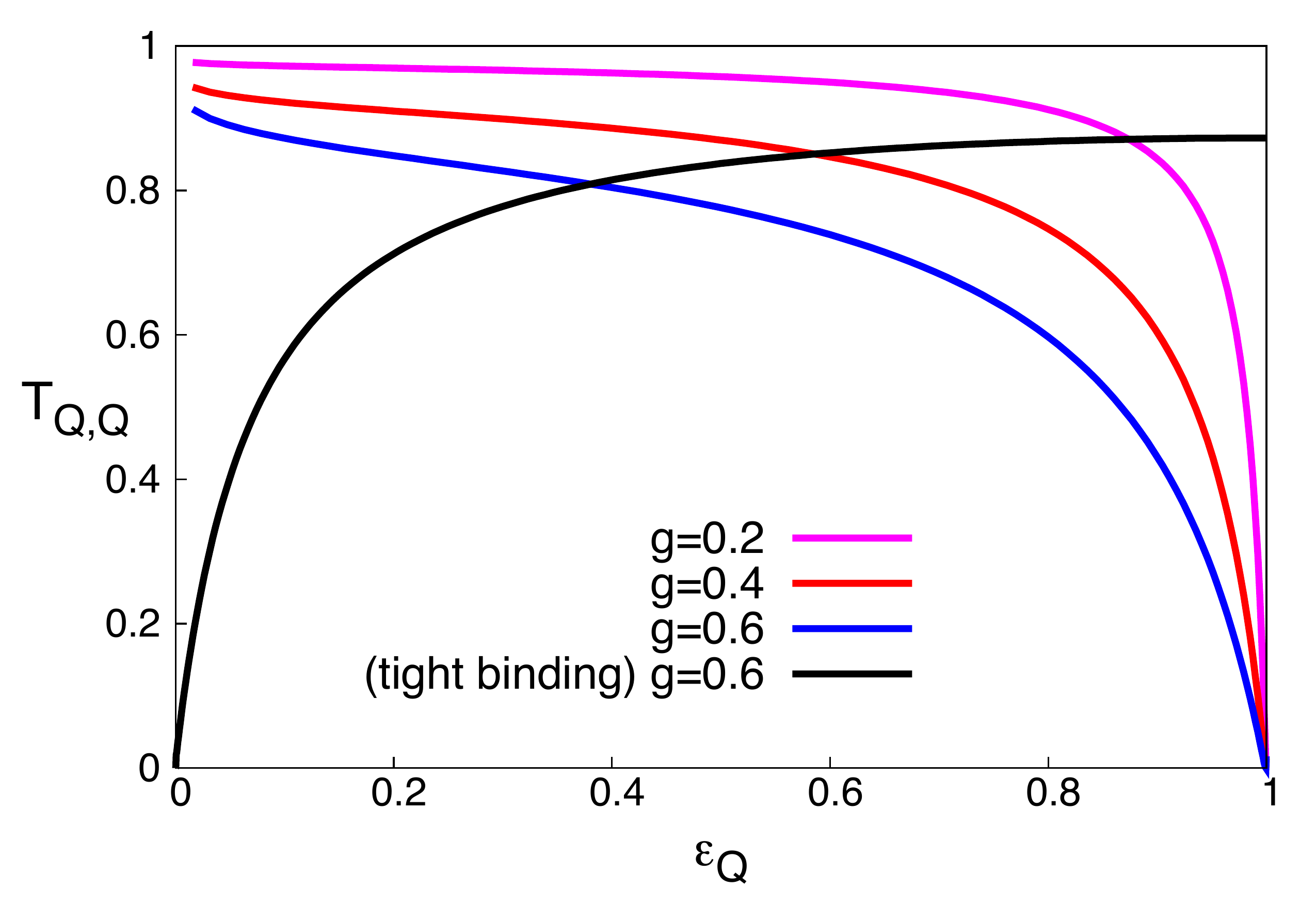}
\caption{${\mathcal T}_{Q,Q}$ vs $\varepsilon_Q$ for a delta-like potential 
of strength $g$ 
for a "free" spinon and a particle in a tight-binding model}
\label{fig6}
\end{figure}

In Fig.\ref{fig6} we show that the 
"free" spinon transmission probability and that of a particle 
in a tight-binding 
model with dispersion relation $\varepsilon_Q=v_s(1-\cos Q)$ behave 
very differently. The "free" spinon transmission probability
is generally a decreasing function of the energy, a property 
of the specific bounded spectrum.
Moreover, we observe that in the linear part of the energy dispersion 
we have high transmission probability which is related  to the fact 
that in a purely linear dispersion relation, i.e. a massless one 
dimensional Dirac equation only a phase is induced in the wavefunction 
and there is no reflection probability \cite{robinson}. 
Additionally from the specific form of the spinon dispersion relation 
we observe that when $\varepsilon_{Q}$ decreases, $u_Q$ increases, 
which implies that $\mathcal{T}_{Q,Q}$ is an increasing function of the spinon 
velocity. Thus a more sensible quantity for the description of the 
transmission coefficient is the spinon velocity and not the spinon energy 
as in usual scattering problems.

\subsection{One-site longitudinal potential}

\begin{figure}
\centering
\includegraphics[width=1.0\linewidth]{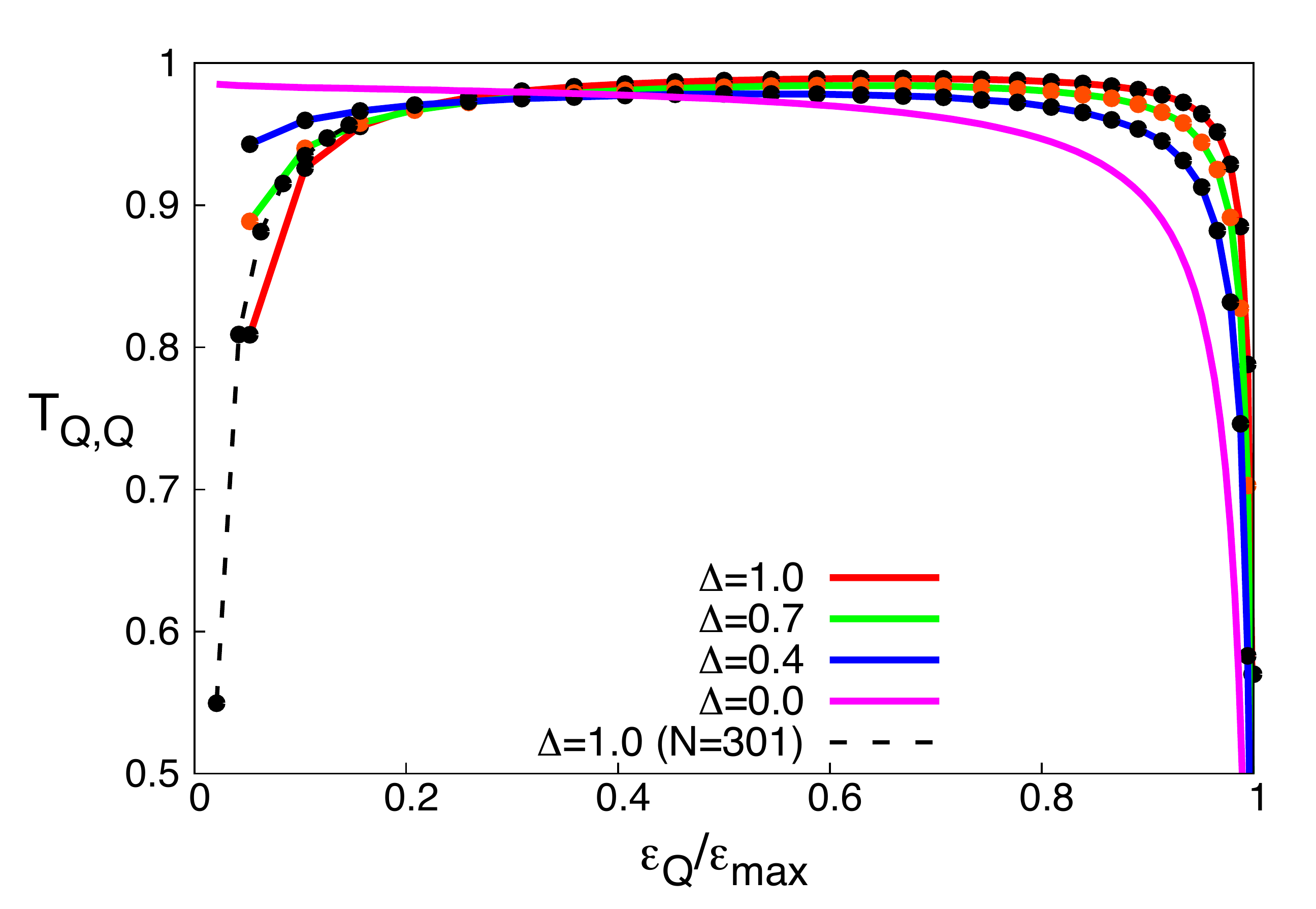}
\caption{${\mathcal T}_{Q,Q}$ vs $\varepsilon_Q$ for various $\Delta$, 
$g=0.15$  for an odd spin chain, $N=121$. 
The black dashed line indicates the $N=301$ data. The solid lines are guides 
to the eye.}
\label{fig7}
\end{figure}

\begin{figure}
\centering
\includegraphics[width=1.0\linewidth]{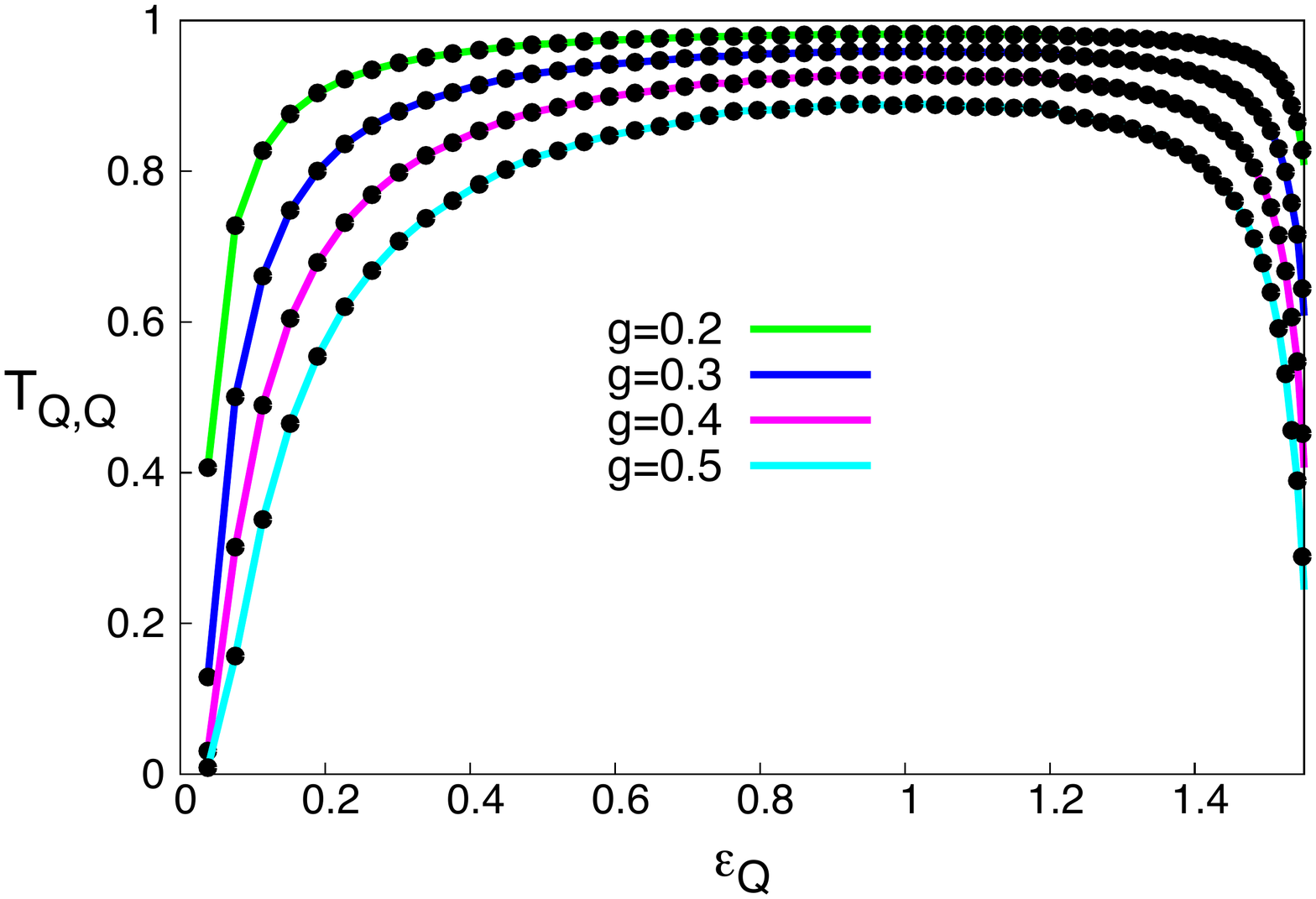}
\caption{${\mathcal T}_{Q,Q}$ vs $\varepsilon_Q$ for 
an odd $N=241$ isotropic model spin chain for various $g$. 
The solid lines are guides to the eye.}
\label{fig8}
\end{figure}

We first consider the scattering of a spinon in an odd-site chain 
from a one-site potential 
$V=g S^z_n$. In the fermionic language of the $t-V$ model \cite{emery} 
this would indeed correspond to the scattering of a spinless fermion from 
a one-site potential. 
In our calculation of the transmission 
coefficient ${\mathcal T}_{Q,Q}$ as a function of spinon energy, 
Fig.\ref{fig7}, we include only the lower one spinon branch 
as intermediate states. 
For $\Delta=0$ we recover the free-spinon result of Fig.\ref{fig6}, 
while for finite $\Delta$ we find a strong suppression of the transmission 
probability at low energies. Because of the finite size of the chain
we cannot study the zero energy limit, however 
we expect the transmission to vanish at this limit as implied by
comparing the $N=121$ and $N=301$ data at low energies.
We should also note that the results are practically independent 
of system size, at least in this lowest branch approximation. 
Similar results are shown in Fig.\ref{fig8} 
for the isotropic model at different potential strengths $g$ 
where, as expected, the transmission is suppressed with increasing 
potential strength. Furthermore, as in the free spinon case, note the 
vanishing of the transmission at high energies, related to the zero
spinon velocity at the top of the energy dispersion.

\begin{figure}[h!]
\centering
\includegraphics[width=1.0\linewidth]{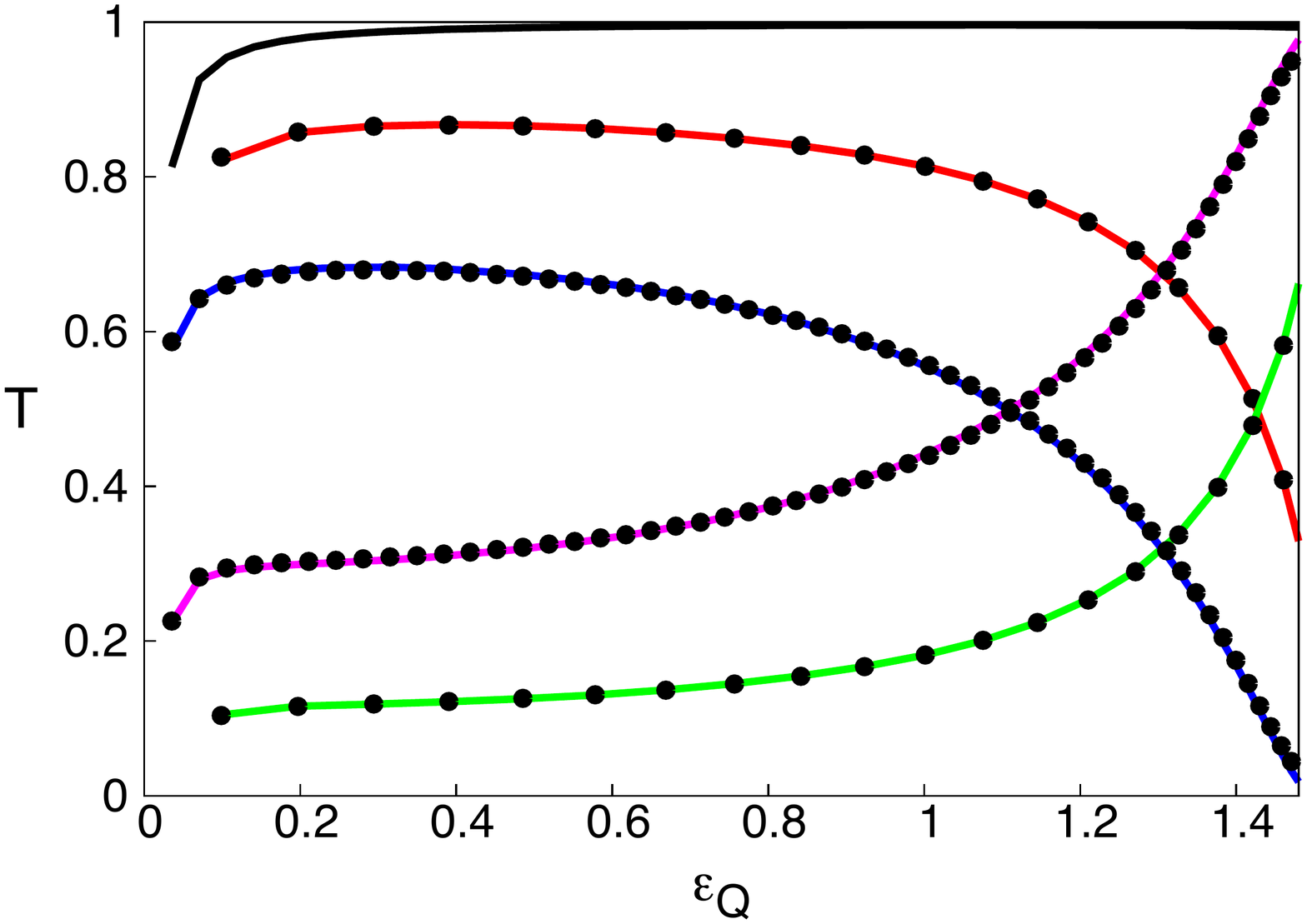}
\caption{ 
${\cal T}_{Q,Q}$ (red $N=100$, blue $N=280$) and ${\cal T}_{Q,Q+\pi}$ 
(green $N=100$, purple $N=280$)  
vs $\varepsilon_{Q}$ for $g=0.15$ and $\Delta=1$. The sum 
${\cal T}_{tot}={\cal T}_{Q,Q}+{\cal T}_{Q,Q+\pi}$ for $N=280$ 
is indicated by  a solid black line.
The solid lines represent the analytical results while the 
dots the numerical data. } 
\label{fig9}
\end{figure}

\begin{figure}[h!]
\centering
\includegraphics[width=1.0\linewidth]{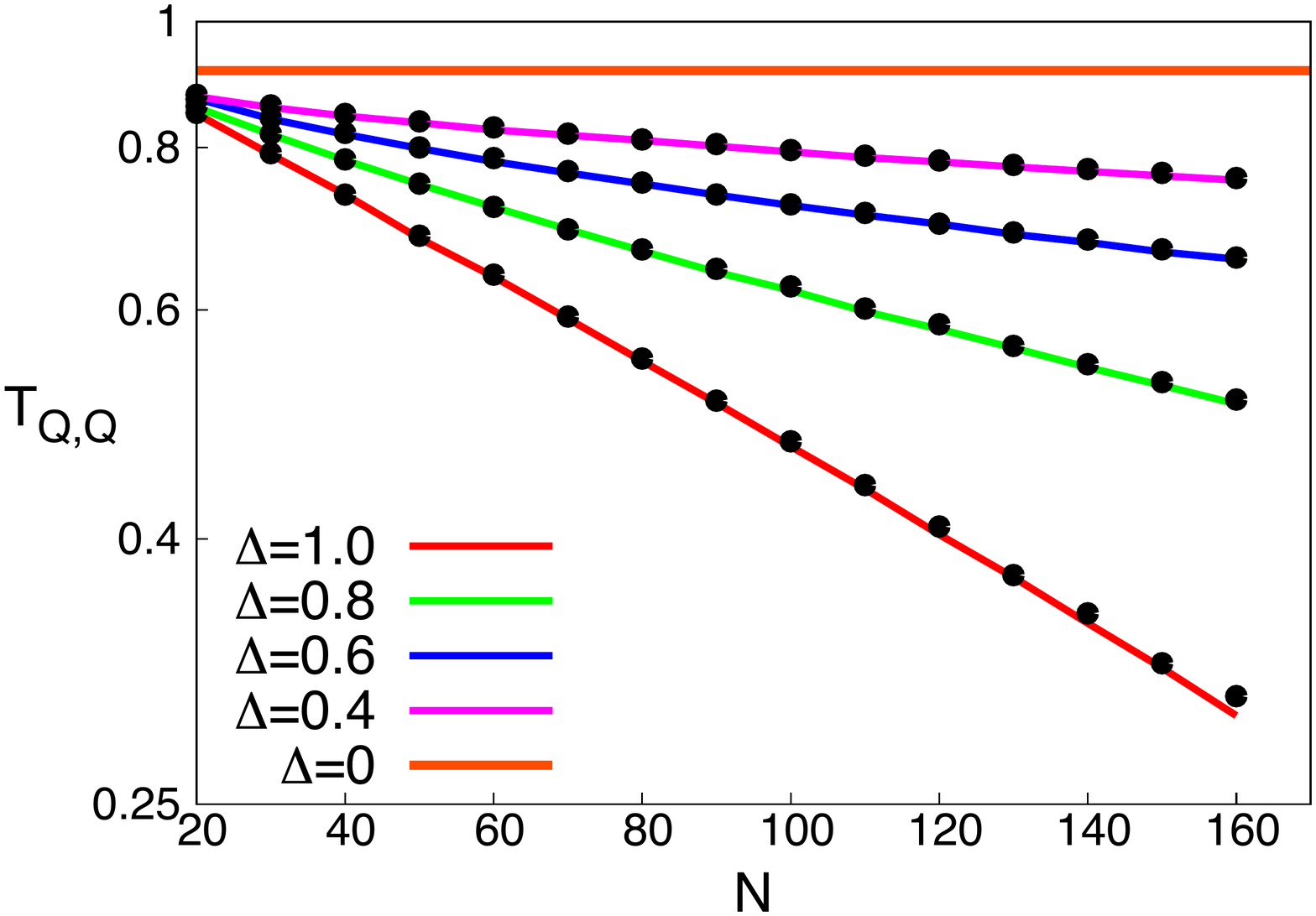}
\caption{ $\log {\cal T_{Q,Q}}$ vs $N$ for $V= g S^z_n$, $g=0.2$ for constant 
energy $\varepsilon_Q/v_s=\sin(2\pi/10)$ and 
$\Delta=0.4,0.6,0.8,1.0$ The solid lines represent the 
analytical approximation \cite{suppl}, while the dots represent 
the numerical data. The horizontal represents the $\Delta=0$ case.} 
\label{fig10}
\end{figure}

\begin{figure}[h!]
\centering
\includegraphics[width=1.0\linewidth]{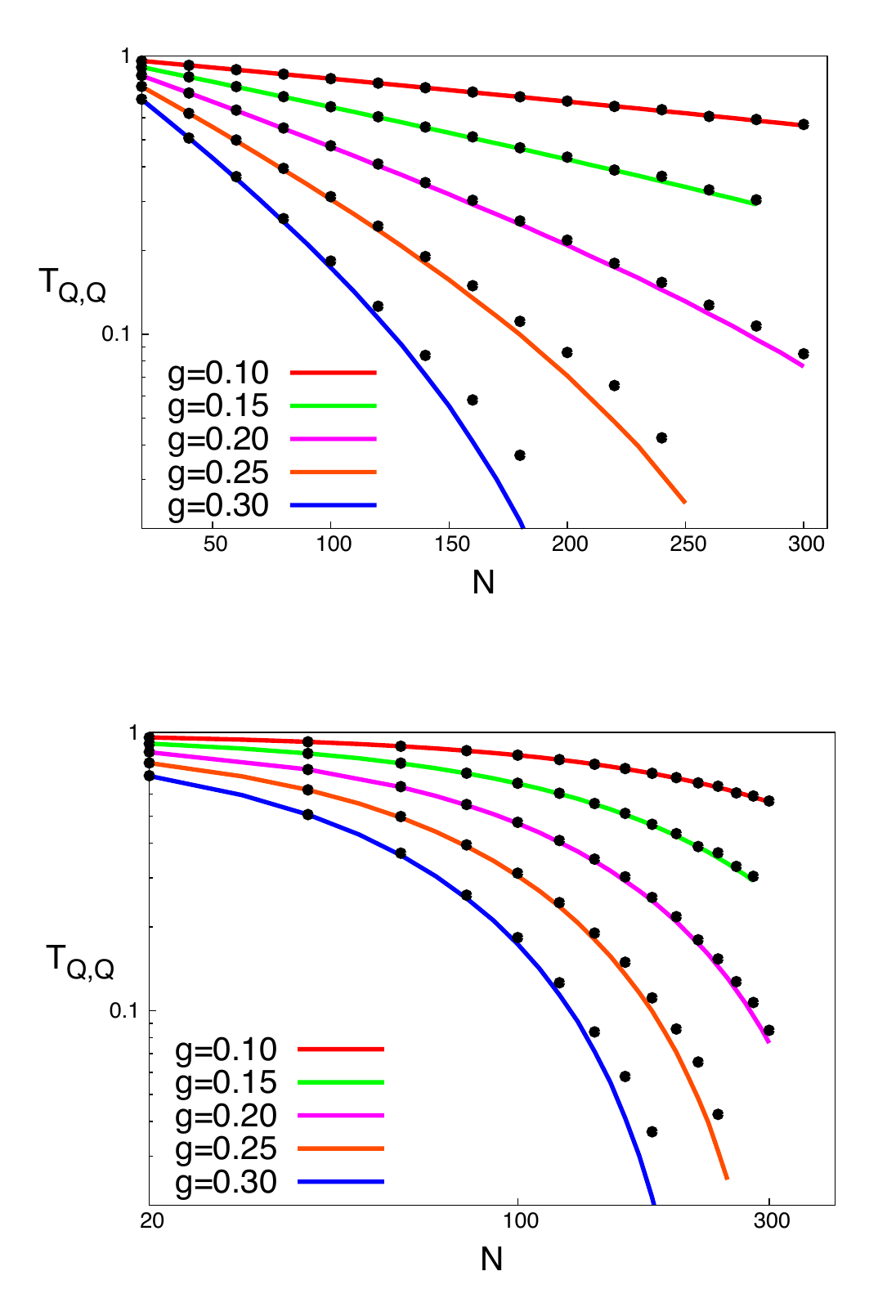}
\caption{ $\log {\cal T}_{Q,Q}$ vs $N$ (top) and  
$\log {\cal T}_{Q,Q}$ vs $\log N$ (bottom) for $V= g S^z_n$ for the 
isotropic model $\Delta=1$ and energy 
$\varepsilon_{Q}={\pi\over 2}\sin(2\pi/10)\simeq 0.92$. 
The solid lines represent the analytical approximation \cite{suppl}, 
while the dots represent the numerical data.}
\label{fig11}
\end{figure}

\begin{figure}[h!]
\centering
\includegraphics[width=1.0\linewidth]{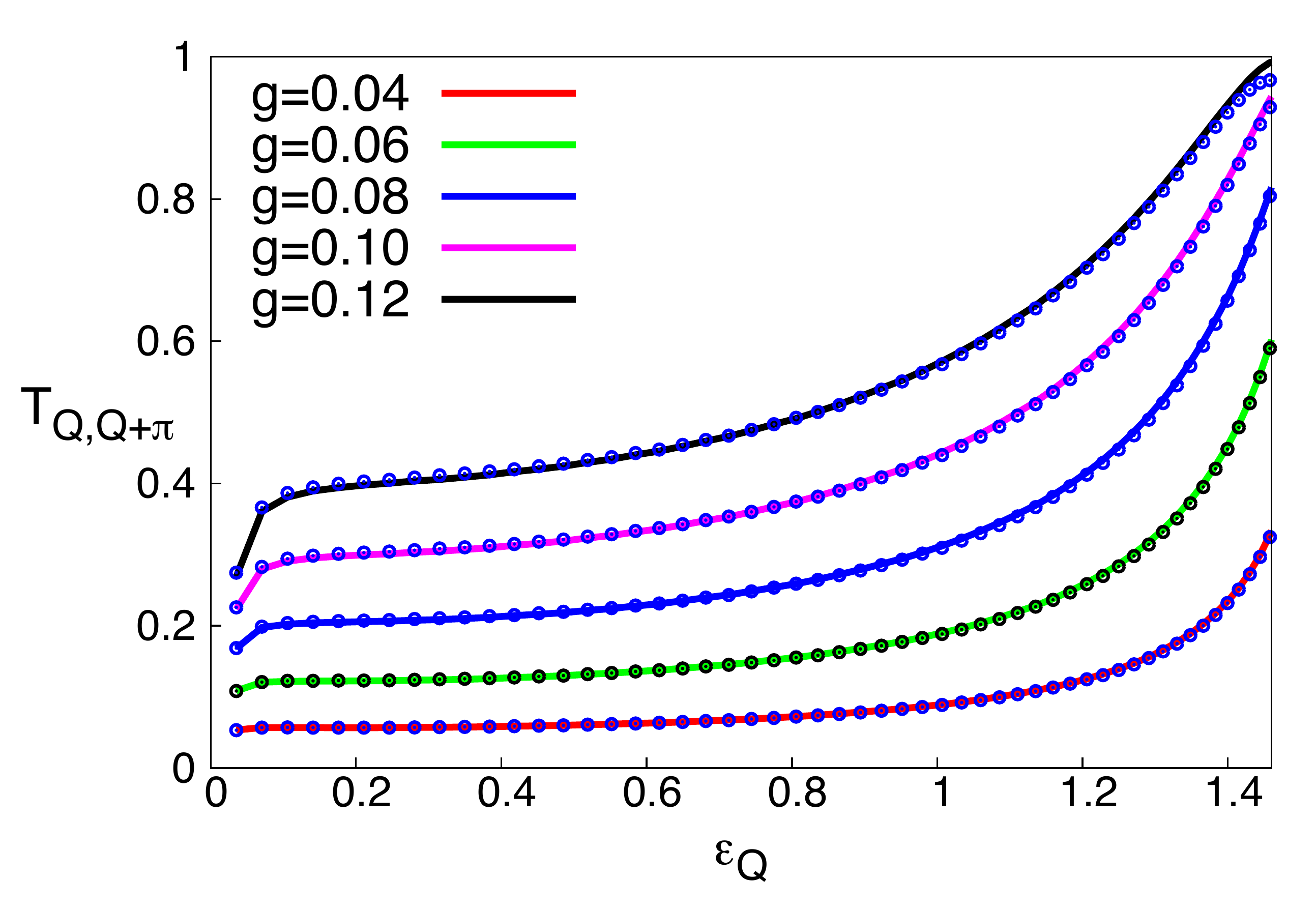}
\caption{ ${\cal T}_{Q,Q+\pi}$ vs $\varepsilon_{Q}$ 
as a function of $g$ for $N=280$. The solid lines are guides to the eye.}
\label{fig12}
\end{figure}

As shown in Fig.\ref{fig3}, in an even chain there are 
two low energy spinon branches. In Fig.\ref{fig9} we find that there 
is a complementarity in transmission, as when ${\mathcal T}_{Q,Q}$ decreases, 
${\mathcal T}_{Q,Q+\pi}$ increases. The sum of the two closely resembles 
the transmission of the one spinon in a odd chain. 
Furthermore, there is a strong size 
dependence of ${\mathcal T}_{Q,Q}$ which can 
probably best be described as exponentially 
decreasing with $N$. 
This is argued in \cite{suppl} and shown in Fig.\ref{fig11} 
where for comparison a power law dependence is also plotted (not shown, 
there is a corresponding exponential increase of ${\mathcal T}_{Q,Q+\pi}$).
The exponential dependence increases with $\Delta$ as shown 
in Fig. \ref{fig10} and with $g$, Fig.\ref{fig11}.
However, the sum 
${\mathcal T}_{Q,Q}+{\mathcal T}_{Q,Q+\pi}$ of transmission probabilities 
shows a 
weak size dependence and of course in the $\Delta=0$ case coincides 
with the one spinon in an odd chain with no size dependence.
In other words, we conjecture that 
in the thermodynamic limit an incoming spinon from the one 
branch is fully transmitted/reflected in the other branch. 
In this calculation we have again included as intermediate states only the 
two lower spinon branches. 
As discussed below, including all the two-spinon 
states, only quantitatively changes this behavior.
Another aspect of this transfer of transmission probability from the 
${\mathcal T}_{Q,Q}$ to the ${\mathcal T}_{Q,Q+\pi}$ branch is shown in 
Fig.\ref{fig12} where we see that ${\mathcal T}_{Q,Q+\pi}$
increases with potential strength. 

Based on the integrable structure of the Heisenberg model
we can understand these results from first principles \cite{suppl}. 
Re-summing to all orders the most important on-shell 
matrix elements, $\vert Q\rangle \rightarrow 
\vert Q+\pi\rangle, \vert \pi-Q\rangle$ 
described in the previous section, 
we obtain a fairly good description of the transmission probabilities
(even quantitative in the weak coupling limit). 
It is easily proved that these transitions  
result in a monotonically decreasing (increasing) transmission 
probability ${\cal T}_{Q,Q}({\cal T}_{Q,\pi+Q})$ 
with spin chain length $N$. 
We expect this behavior to be generic in one dimensional spin chains,
simply here, the integrability of the model allows us to explicitly 
evaluate the corresponding exponents.

\begin{figure}[h!]
\centering
\includegraphics[width=0.80\linewidth]{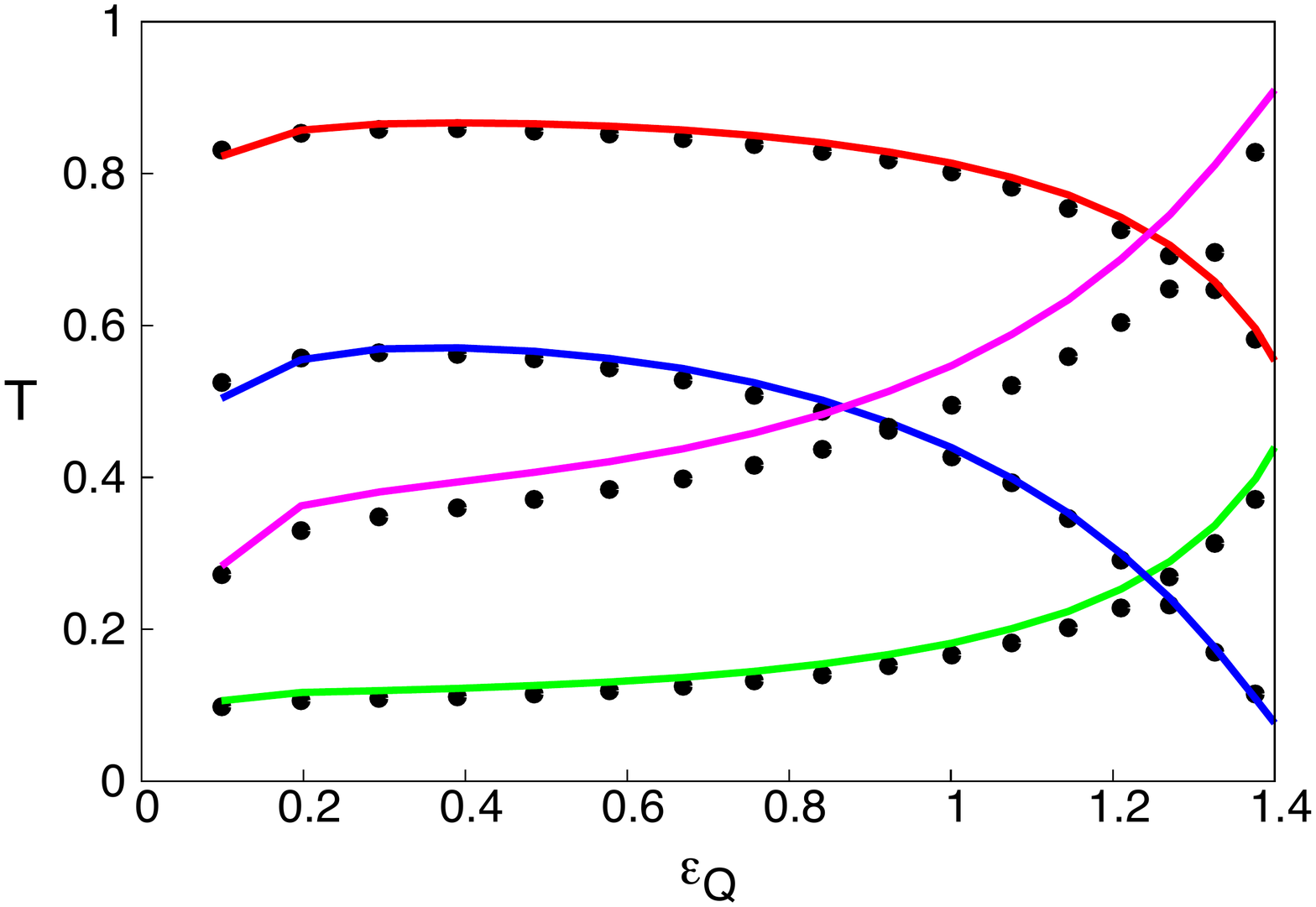}
\caption{${\cal T}_{Q,Q}$ 
(red $g=0.1$, blue $g=0.2$) 
and ${\cal T}_{Q,Q+\pi}$ 
(green $g=0.1$, purple $g=0.2$)  
vs $\varepsilon_{Q}$ for $N=100$ and $\Delta=1$.
The solid lines are produced by including only the lower branch while 
the dots represent the numerical data obtained by including the whole 
two-spinon continuum.} 
\label{fig13}
\end{figure}

\begin{figure}[h!]
\centering
\includegraphics[width=0.80\linewidth]{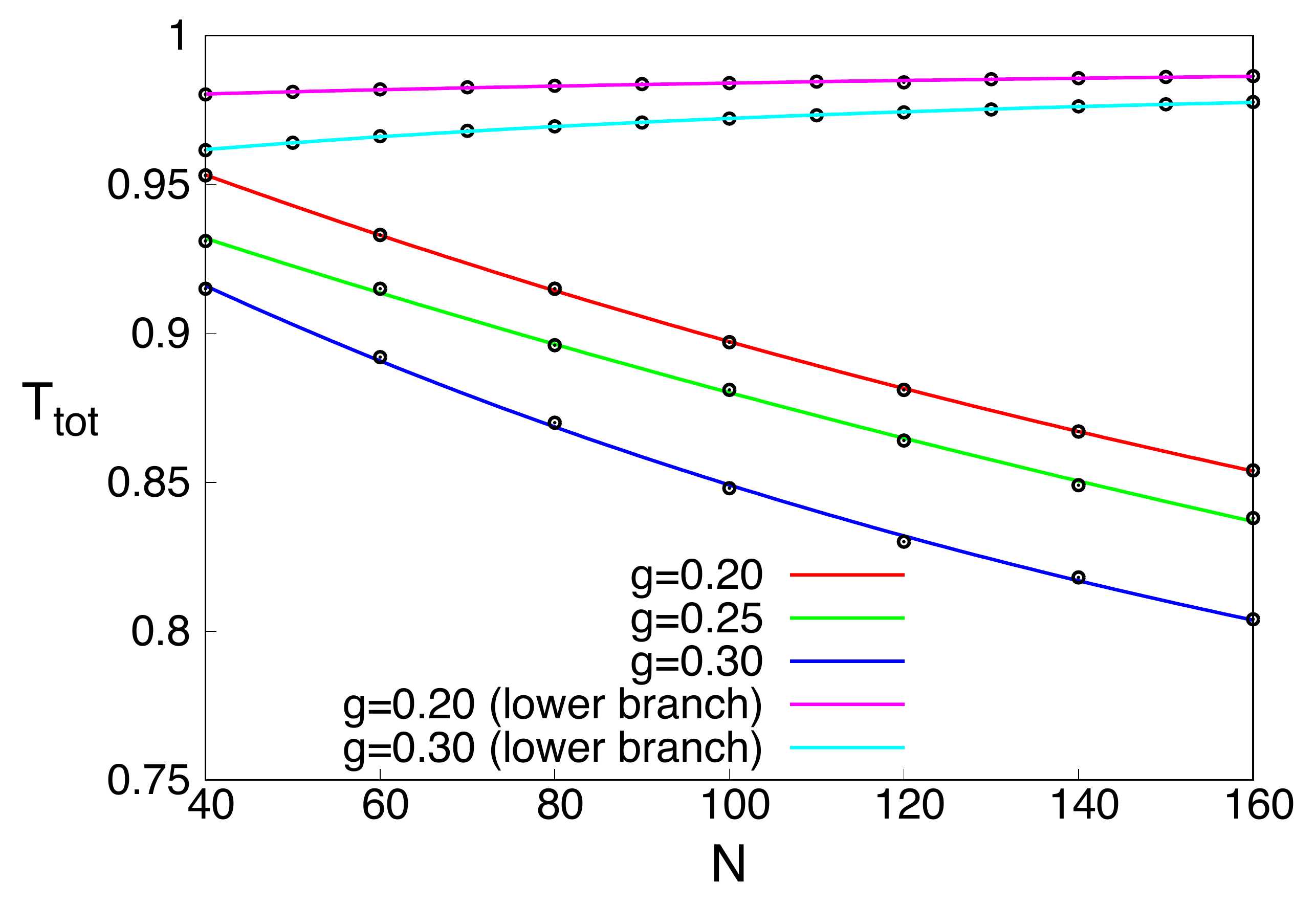}
\caption{$\mathcal{T}_{tot}$ versus $N$ for the $V=gS^z_n$ potential 
$\Delta=1$ and $\varepsilon_Q \simeq 0.923$ 
including the two-spinon continuum along the lower branch data. 
The dots represent 
the numerical data, while the solid lines represent the fitted curve 
$\mathcal{T}_{tot}=A\exp(-B f^z(Q) g^2N/u^2_Q)+C$.} 
\label{fig14}
\end{figure}

By a numerical fit in Figs.\ref{fig10},\ref{fig11} 
we find that a useful quantity for the 
description of the scattering process is $ g_{eff}=g N^{1-\mathcal{Z}^2}$ 
and that for $\varepsilon_Q$ not close to zero the transmission 
coefficient behaves as 
\be 
{\mathcal T}_{Q,Q} \simeq e^{-a (g_{eff}/u_Q)^2}
\label{eq4}
\ee
which holds for $g_{eff}/u_Q<<1$. 
Thus for the isotropic Heisenberg model ($\gamma=0,~\Delta=1$) 
which is the most experimentally relevant 
$\mathcal T_{Q,Q} \simeq  e^{-a (g/u_Q)^2 N}$.
Although this approach does not offer an 
analytical solution of the scattering problem, using the framework 
of integrability we derived a connection between the transmission 
coefficients and $\theta_{zz}=2\mathcal{Z}^2$, the critical exponent 
of the ground state's correlation function 
$\langle 0|s^z_1 s^z_{n+1}|0\rangle$  dominant oscillatory part. 
Predicted by CFT and Bethe Ansatz 
calculations \cite{korepin,kitanine}, it offers a qualitative description 
of the scattering process.

Note that this approximation gives reasonable results 
even though we have performed a rough elimination 
of most of the intermediate matrix elements. 
On the other hand from the specific form of 
the transmission probability of the  "free" spinon model we observe 
that the dominant behavior is given by the on-shell matrix elements 
and the rest of the matrix $V$ serves as a correction, 
which justifies the reasoning for the above approximation. 
Of course, as we see in Fig.\ref{fig11}, it is a weak coupling 
approximation, albeit a very good one, that becomes increasingly 
unreliable in the strong $g$ coupling limit.
Even more, in the strong coupling $g/u_Q >> 1$ limit 
(e.g. $Q\rightarrow \pm \pi/2)$ the numerical 
T-matrix approach we are using often does not converge at all. 

Finally, to improve on the lower branch approximation we include 
all the two-spinon Cloiseaux-Pearson \cite{cloizeaux} 
states which forces us however to study rather 
small size spin chains as the space of intermediate states increases as $N^2$. 
As shown in Figs. \ref{fig13} and \ref{fig14}
it results to only quantitative differences with most of the effect 
coming from the ${\cal T}_{Q,Q+\pi}$ transition. We estimate the 
$N\rightarrow +\infty$ extrapolated value of ${\cal T}_{tot}$ 
to be reduced by about 
30 percent from the value of the lower branch data. We should note 
however that the two-spinon continuum data become increasingly sensitive 
with system size to details of the calculation e.g. separation 
of the real and imaginary part in the T-matrix numerical evaluation.

Comparing the even and odd site case, 
we found an interesting topological effect.
In the odd chains, in our one-spinon study where the spectrum 
is two-fold degenerate, we find a rather regular behavior 
of scattering coefficients. In the even chains, due to the topological 
two-spinon constraint, we have a four-fold degenerate spectrum 
that, together with the singular $\pi-$transition, implies a transfer  
of transmission probability between the two spinon branches.
Thus, in the spinon scattering, we have an interplay of the 
topological character and the singular matrix elements of a critical system.

\subsection{Spin-phonon potential}

The spin-phonon interaction is 
described by a one-link potential of the form 
\be
V=g(S^{-}_nS^{+}_{n+1}+S^{+}_nS^{-}_{n+1}).
\ee

\begin{figure}[h!]
\centering
\includegraphics[width=1.0\linewidth]{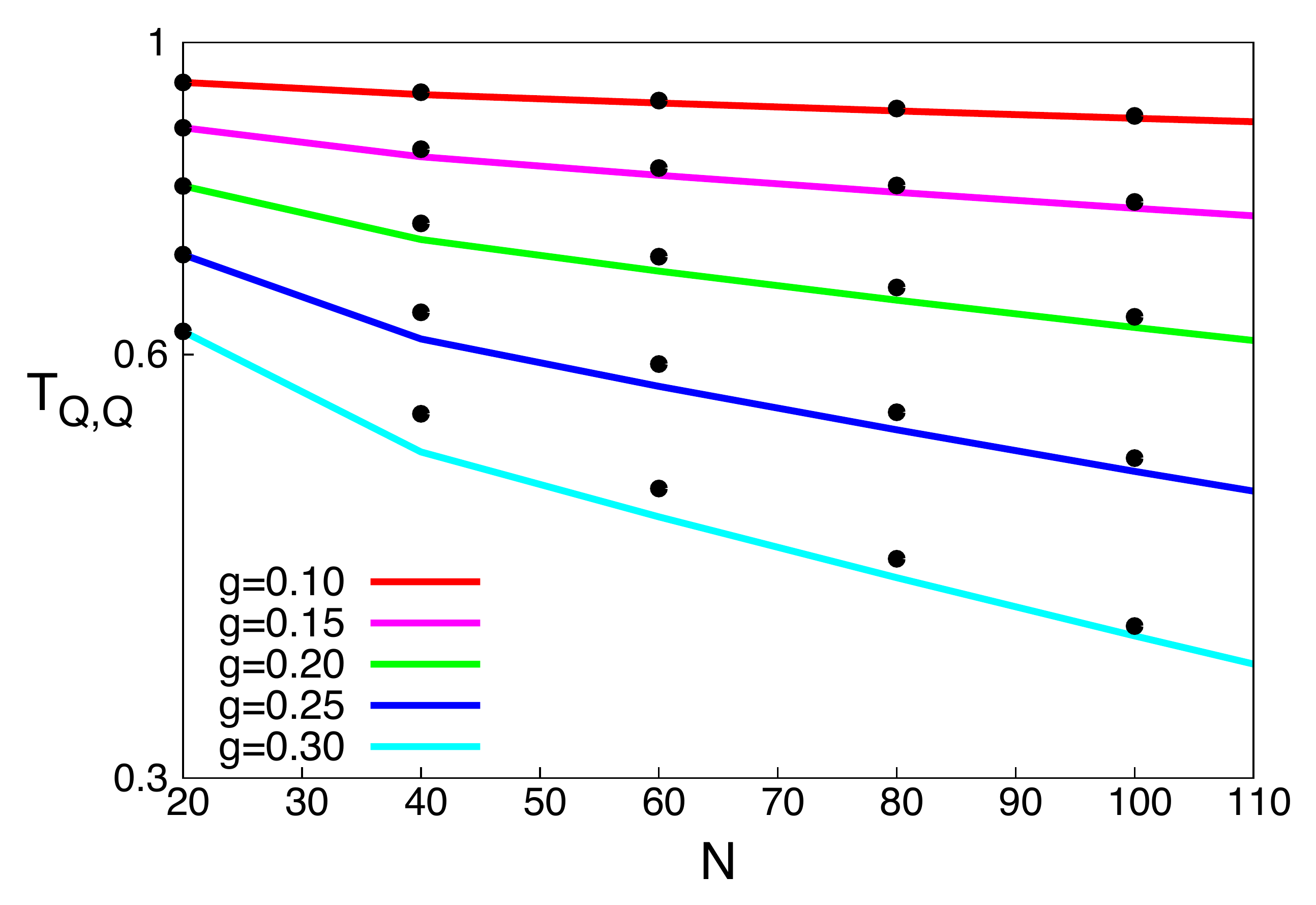}
\caption{$\log \mathcal {T}_{Q,Q}$  vs $N$ for a one-link spin-phonon 
potential $V$, $\varepsilon_{Q}\simeq 0.92$ 
and $g=0.1,0.15,0.2,0.25,0.3$. Solid lines prediction based on the dominant 
matrix elements \cite{suppl}.} 
\label{fig15}
\end{figure}

In Fig.\ref{fig15} the numerical calculation for an even-site chain shows 
that $\mathcal{T}_{Q,Q}\rightarrow 0$ 
as $N$ increases. Similarly to the previous case, we obtain an approximate 
analytical result by using the dominant matrix elements that were described 
in the previous section. In particular, the monotonicity of the scaling factors 
implies that the transmission and reflection coefficients will be scale 
invariant for $\Delta=0$ while on the contrary,  for $ 0 < \Delta \le 1$
$\mathcal{T}_{Q,Q}\rightarrow 0$ as $N$ increases. 
Moreover, the relation of the scattering coefficients to the spinon 
energy $\varepsilon_{Q}$ is very similar to that of a longitudinal magnetic 
potential as was depicted in Fig.\ref{fig9}.

\subsection{Transverse potential}
We now turn to a transverse magnetic potential, $V=g S^x_n$. The main 
difference of this potential to the two previous ones we studied  
is that it acts non trivially only between states with $\Delta S^z=\pm 1$. 
We will restrict ourselves to transitions between the $S^z=1$ and $S^z=2$ 
magnetization sectors.

\begin{figure}[h!]
\centering
\includegraphics[width=1.0\linewidth]{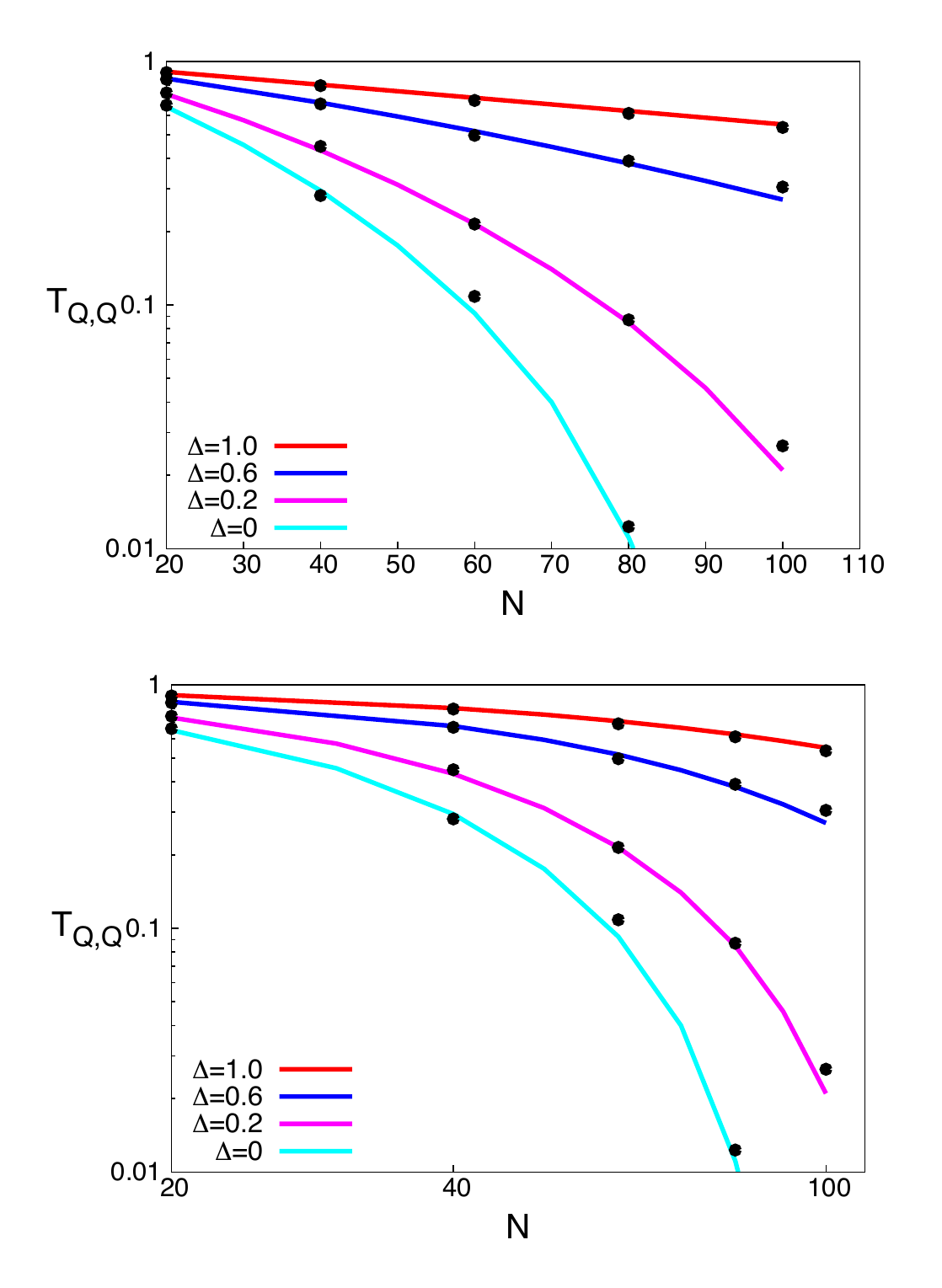}
\caption{$\log \mathcal{T}_{Q,Q}$ vs system size $N$ (top) 
and $\log \mathcal{T}_{Q,Q}$ vs system size $\log N$  for a one-site 
transverse potential $V=g S^x_n$, $g=0.2$, 
$\varepsilon_{Q}=v_s\sin ({2\pi\over 10})$ 
and $\Delta=1,0.6,0.2,0.0$. Solid lines are the prediction considering 
the dominant matrix elements \cite{suppl}.}
\label{fig16}
\end{figure}

Fig.\ref{fig16} shows that similarly to the previous cases,
for an even-site chain  
$\mathcal{T}_{Q,Q}\rightarrow 0$ as $N$ increases. 
Again the dependence is probably best be described as exponential, 
as argued in \cite{suppl} and by comparison with a power law one.
However, this time we find 
that this holds also for $\Delta=0$ and in fact the scattering 
increases as $\Delta$ decreases, which is the opposite to what happened 
in the previous cases. Again we can obtain a qualitative explanation of this 
behavior by using the fact \cite{kitanine1,kitanine2} that the dominant 
matrix element approximately scales as 
$\theta_{-+}={1\over 2\mathcal{Z}^2}\simeq {\pi-\gamma\over \pi}$ 
which is the dominant critical exponent of the ground state correlation 
$\langle 0|\sigma^{-}_1\sigma^{+}_{n+1}|0\rangle$. 
By re-summation \cite{suppl} and the monotonicity of the 
critical exponents with respect to $\Delta$ one can argue that 
$\mathcal{T}_{Q,Q} \rightarrow 0$ for $0 < \Delta \le 1$. 
Nevertheless, a full scale 
analysis of the matrix elements should be done in order to give a 
definite answer. Similarly to the previous cases, by defining  
$g_{eff}\equiv gN^{1-{1\over 4\mathcal{Z}^2}}$ implying
$g_{eff}=g \sqrt{N}$ for $\Delta=1$ (isotropic model) 
and $g_{eff}=g N^{3/4}$ for $\Delta=0$ (XY model), we conclude that
\begin{equation}
\mathcal{T}_{Q,Q}\simeq e^{-f^x(Q) (g_{eff}/u_Q)^2}
\label{tspdom}
\end{equation}
\noindent
in the region $g_{eff}/(4 u^2_Q/f^x(Q))<<1$, a behavior 
which agrees well with the numerical data.

\subsection{Extended potential}
\begin{figure}[h!]
\centering
\includegraphics[width=1.0\linewidth]{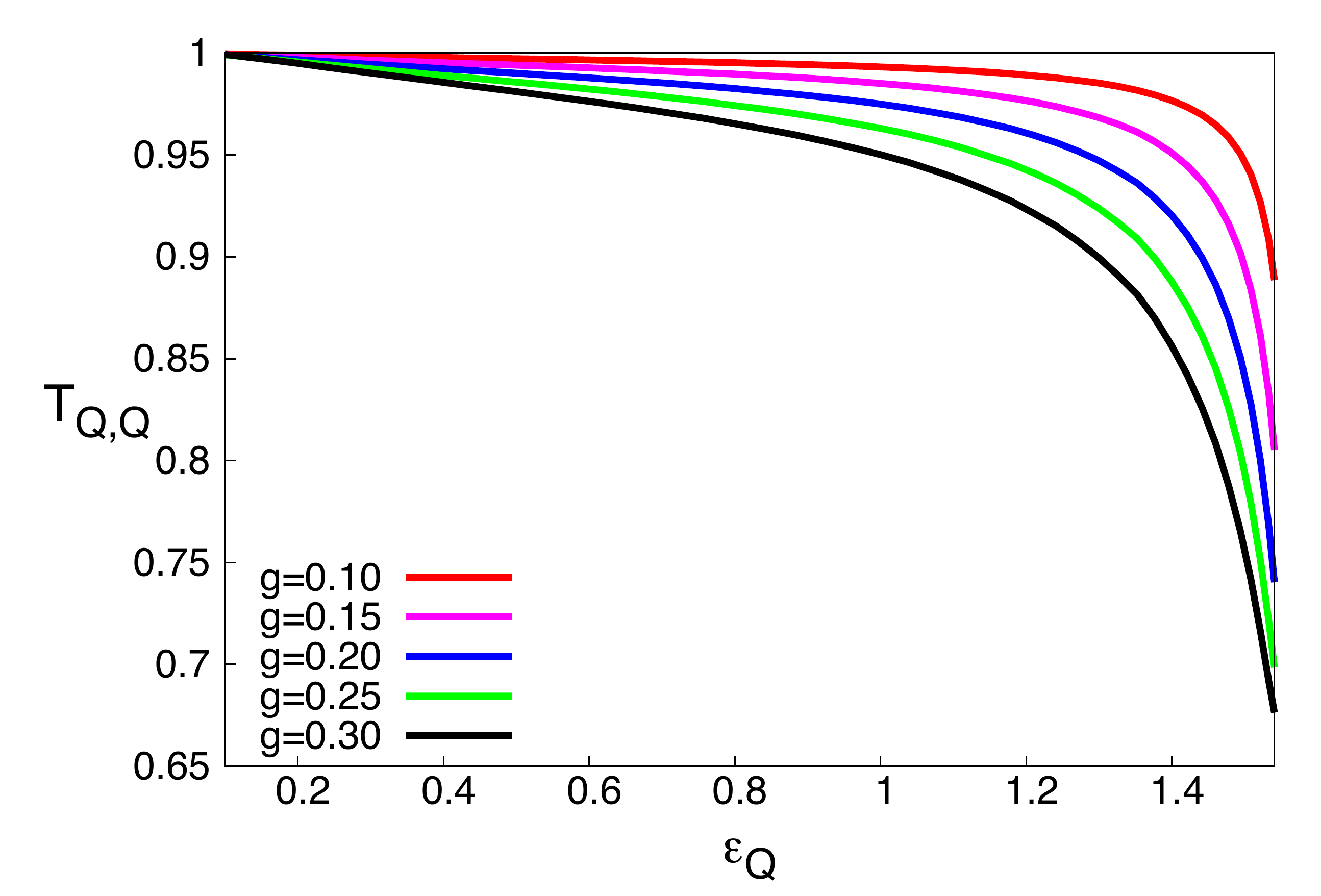}
\caption{ $\mathcal{T}_{Q,Q}$ vs spinon energy $\varepsilon_{Q}$ for a 
potential $V=g (S^z_n+S^z_{n+1})$, $N=201$. } 
\label{fig17}
\end{figure}

Finally, we consider the spinon scattering from an extended potential
\be 
V_{ext}=\sum_{n=1}^m g_n V_n.
\ee
By the numerical procedure presented earlier,
we can calculate the transmission 
probability for an arbitrary potential profile $\lbrace g_n\rbrace$ 
in the two-spinon continuum approximation. 
We start with the scattering of a spinon in an odd chain by 
a two-site longitudinal potential, a case analogous to Fig.\ref{fig7} for a 
on-site potential. In Fig.\ref{fig17} we see a remarkable difference 
at low energies where there is complete transmission. 
This situation is consistent with the well known "cutting" and "healing" 
\cite{kane,affleck} effect in one dimensional correlated systems and spin 
chains, where one weak-link is cutting a chain at low energies while two
weak-links are healed. This effect leads to a finite conductance  
with a power law dependence on the temperature due to thermal 
effects. 

Here, we can understand the results of extended potentials by considering
the "diffraction" relation (\ref{vm}). For an $m=2$ longitudinal potential 
in an odd chain, at low energies $Q\rightarrow 0$, the $q=\pi-2Q$ 
scattering matrix element vanishes leading to total 
transmission. Following the same argument, we also find that for an even 
chain with an $m=2$ longitudinal potential the transfer 
of transmission probability
from ${\cal T}_{Q,Q}$ to ${\cal T}_{Q,Q+\pi}$ found in Fig.\ref{fig7} 
is now totally suppressed as the $q=\pi$ matrix element vanishes.
Following the same line of re-summation of dominant matrix elements and by  
taking into account the corresponding "diffraction" factor allows us to 
understand the transmission by extended potentials.

\section{Conclusions}
Using the Bethe ansatz method and the T-matrix approach, 
we have studied the scattering of a spinon from prototype potentials.
Three main features emerged from this study;
first, we are considering a quantum many-body problem, so in principle 
outgoing states with creation of spinons or "electron-hole" pairs 
is possible although we expect from the scattering matrix elements 
that these processes have lower probability.
We have limited our study to outgoing states with the same number 
of spinons as the incoming state.
Second, we can qualitatively 
account for the transmission probabilities by re-summing 
the dominant scattering elements. Their dependence on the size of the 
spin chain is given by the critical exponents characterizing the 
anisotropic Heisenberg model. 
Thus, we linked the scattering to the critical properties 
of this integrable model and we evaluated them by the 
Bethe ansatz method. It is an open, technically very difficult, 
question whether including all intermediate states $O(2^N)$ 
would qualitatively change the present picture. 
Third, we have found an intriguing topological effect as, in an even chain 
there is complete transfer of the incoming spinon transmission probability
from the one branch of the dispersion  
to the other branch. At the moment, in a macroscopic open chain 
the role of this odd-even effect is ambiguous. Further study 
is necessary to clarify it, presumably including 
further outgoing states, e.g. three-spinon states in odd chains.
Note that, several experimental and theoretical studies \cite{evenodd} 
have addressed the physical effect of even vs. odd chain length 
in the thermodynamic properties of finite size chains.

Along the line of dominant matrix elements, 
we analyzed a basic difference in the scattering 
coefficients of longitudinal
and weak link potentials to those of a transverse potential. 
We also discussed extended potentials where, a drastic dependence 
of scattering coefficients on the potential extent,  
we  attributed it to a geometric "diffraction" factor and dominant 
scattering matrix elements. These results are  
consistent with previous studies on "cutting-healing" 
in 1D correlated systems \cite{kane,affleck,metavitsiadis}.

Considering experiment, we studied the problem of a spinon excited 
above the ground state and scattering from a potential.
Although we have not addressed any particular experiment, our study 
would provide key elements in the interpretation of far-out of 
equilibrium experiments as well as thermal transport ones. 
For instance zero temperature tunneling 
studied by a "Landauer" type approach  
or spinon transport probed e.g. by 
terahertz 2D coherent spectroscopy \cite{thz} experiments. 

\section{Acknowledgments}
This work was supported by the European Union Program No. 
FP7-REGPOT-2012-2013-1 under Grant No. 316165, the Deutsche 
Forschungsgemeinschaft through Grant HE3439/13, 
the Alexander von Humboldt Foundation, 
the Hellenic Foundation for Research and Innovation (HFRI) 
and the General Secretariat for Research and Technology (GSRT), 
under the HFRI PhD Fellowship grant KA4819. A.P. acknowledges valuable 
discussions with P. Lambropoulos, N. Kitanine and J.S. Caux. 
X.Z. acknowledges fruitful discussions with H. Tsunetsugu, 
A. Kl\"umper, T. Tomaras, C. Hess, B. B\"uchner, 
A. Chernyshev, S. White and the hospitality of the 
Institute for Solid State Physics - U. Tokyo and University of Irvine.

\widetext 

\section{Supplementary Material:
	Scattering of spinon excitations by potentials
	in the 1D Heisenberg model }

\subsection{Dominant Matrix elements }

In this section we briefly describe the prescription for the calculation 
of the matrix elements of the anisotropic Heisenberg spin chain, 
the dominant matrix elements in the scattering processes and their 
relation to the critical exponents that appear in the correlation 
functions.

We begin by using the fact that in the XXZ spin chain, the total 
magnetization commutes with the Hamiltonian, which lead us to partition 
the Hilbert space into subspaces of fixed magnetization, determined from 
the number of reversed spins $M$. For simplicity we take an even number 
of spin sites and $2M\leq N$. Therefore all the eigenstates in each subspace 
are fully described by a set of rapidities $\lbrace\lambda_k\rbrace_{k=1}^M$, which correspond to solutions of the Bethe equations \cite{bethe,book}.  
\be 
\arctan\left[\tanh(\lambda_k)\over\tan(\gamma/2)  \right]-
{1\over N}\sum_{j=1}^M 
\arctan\left[\tanh(\lambda_j-\lambda_k)\over\tan(\gamma)  \right]
={\pi\over N}I_k,
\label{eq1}
\ee
where $\gamma=\arccos \Delta$. $\lbrace I_k\rbrace_{k=1}^M$ are different 
integers or half-integers, defined modulo $N$, which are the analog of 
quantum numbers due to the fact that each particular choice of a set $\lbrace I_j\rbrace_{k=1}^M$ uniquely specifies a Bethe eigenstate. Note that the ground state is given by the configuration $\lbrace {-{M+1\over 2}}+k\rbrace_{k=1}^M$.

The energy and momentum of a state parametrized by the set of rapidities 
$\lbrace \lambda_k\rbrace_{k=1}^M$ or equivalently the set of 
quantum numbers  $\lbrace I_k\rbrace_{k=1}^M$ are given by,
\be 
E_{\lbrace\lambda\rbrace}=
J\sum_{k=1}^M {-\sin^2\gamma\over \cosh 2\lambda_k-\cos\gamma}-h({N\over 2}-M)
\label{eq2}
\ee

\be 
Q_{\lbrace\lambda\rbrace}=\sum_{k=1}^M i\ln {\sinh(\lambda_k+i\gamma/2)\over 
	\sinh(\lambda_k-i\gamma/2)}=\pi M+{2\pi\over N}\sum_{k=1}^M I_k. 
\label{eq3}
\ee

From now on, we redefine $E_{\lbrace\lambda\rbrace}$, 
$Q_{\lbrace\lambda\rbrace}$ to be above the lowest spinon energy and 
momentum respectively and those belonging to the lowest branch of the 
two spinon continuum we write them as $\varepsilon_{Q}$ and $Q$ respectively. 
For $0<\Delta\leq 1$ the two spinon spectrum is defined as the set of 
all the real solutions, with dimension $N(N+2)/8$. 

Kitanine et al. \cite{kitanine,caux}, via an algebraic procedure, calculated 
the form factors for the anisotropic spin chain and successfully reduced 
complicated matrix elements between Bethe states for local spin operators 
to determinant expressions. The matrix elements for the longitudinal, 
transverse and spin-phonon potential are given by the expression 
$\langle \lbrace \lambda\rbrace\vert V_n\vert \lbrace \mu\rbrace\rangle=
{F^{a}_n(\lbrace \lambda\rbrace,\lbrace \mu\rbrace)\over 
	\sqrt{\mathcal{N}(\lbrace \lambda\rbrace)\mathcal{N}(\lbrace \mu\rbrace)}}$, 
where $a=z,x,ph$, $F^{a}_n(\lbrace \lambda\rbrace,\lbrace \mu\rbrace)$ 
is the form factor and $\mathcal{N}(\lbrace\lambda\rbrace)=
\langle \lbrace \lambda\rbrace\vert\lbrace \lambda\rbrace\rangle $ 
the norm of the unnormalized Bethe state 
$\vert \lbrace\lambda\rbrace\rangle$ as fully described in \cite{kitanine}. 
Using these expressions we calculate all the 
matrix elements of the potential matrix written in the Bethe Ansatz basis,
\be
V=\sum_{\lbrace\lambda\rbrace,\lbrace\mu\rbrace} 
\langle \lbrace \lambda\rbrace\vert V \vert \lbrace\mu\rbrace\rangle 
\vert\lbrace\lambda\rbrace\rangle\langle\lbrace\mu\rbrace\vert,
\label{eq3}
\ee
corresponding to longitudinal, transverse and spin-phonon potentials.

\subsubsection{Longitudinal potential $V=g S^z_n$ }

\begin{figure}[h!]
	\centering
	\includegraphics[width=0.8\linewidth]{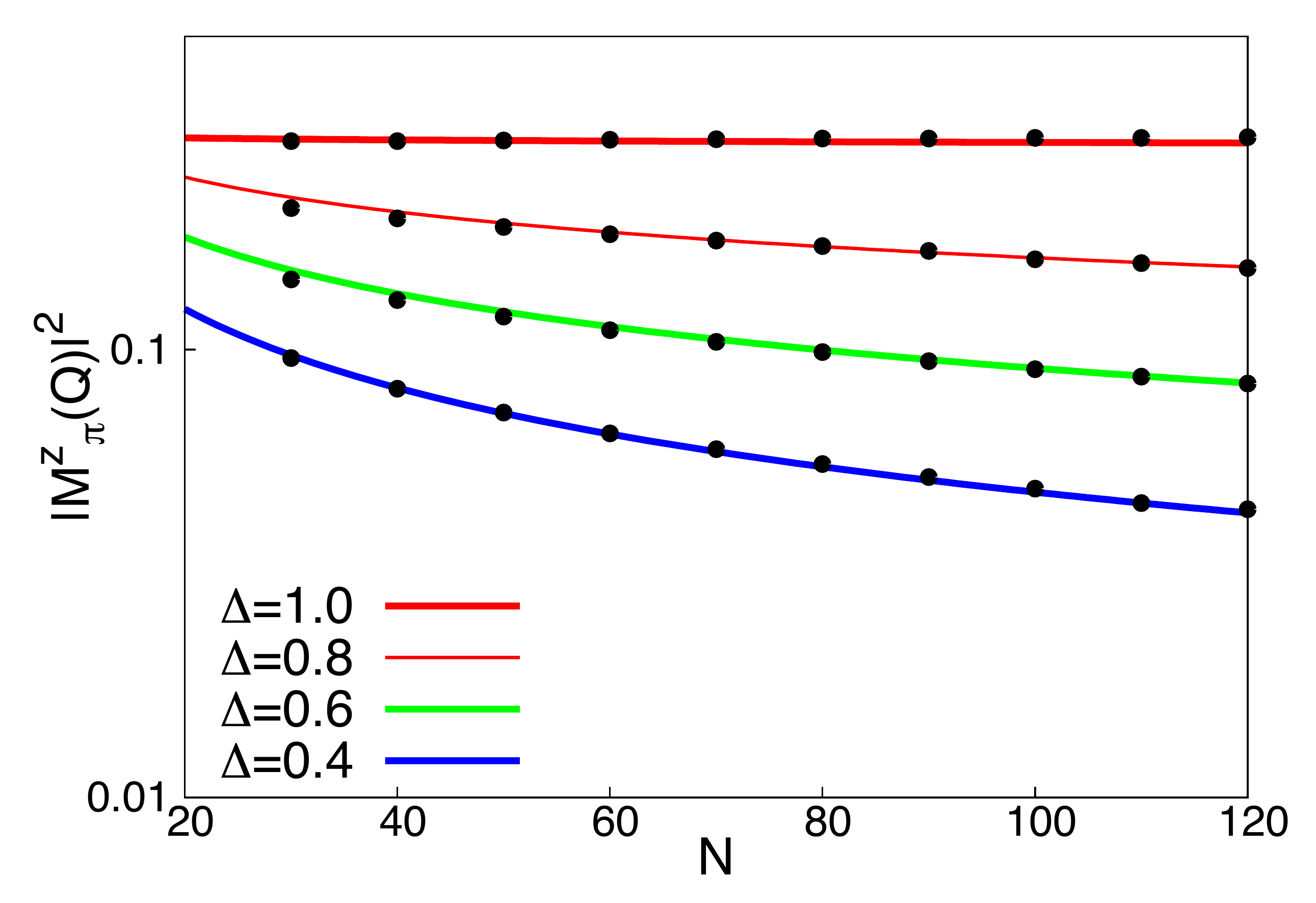}
	\caption{$\vert \langle Q+\pi\vert S^z_\pi\vert Q\rangle\vert^2$ 
		versus $N$ for $\Delta=1,0.8,0.6,0.4$, for $Q=2\pi/10$. 
		The solid lines correspond to the
		fitted curve of the form $\vert \langle Q+\pi\vert S^z_\pi\vert 
		Q\rangle\vert^2\simeq 1/N^{2\mathcal{Z}^2-1}$ while the dots are the 
		numerical data} 
	\label{fig1s}
\end{figure}

\begin{figure}[h!]
	\centering
	\includegraphics[width=0.8\linewidth]{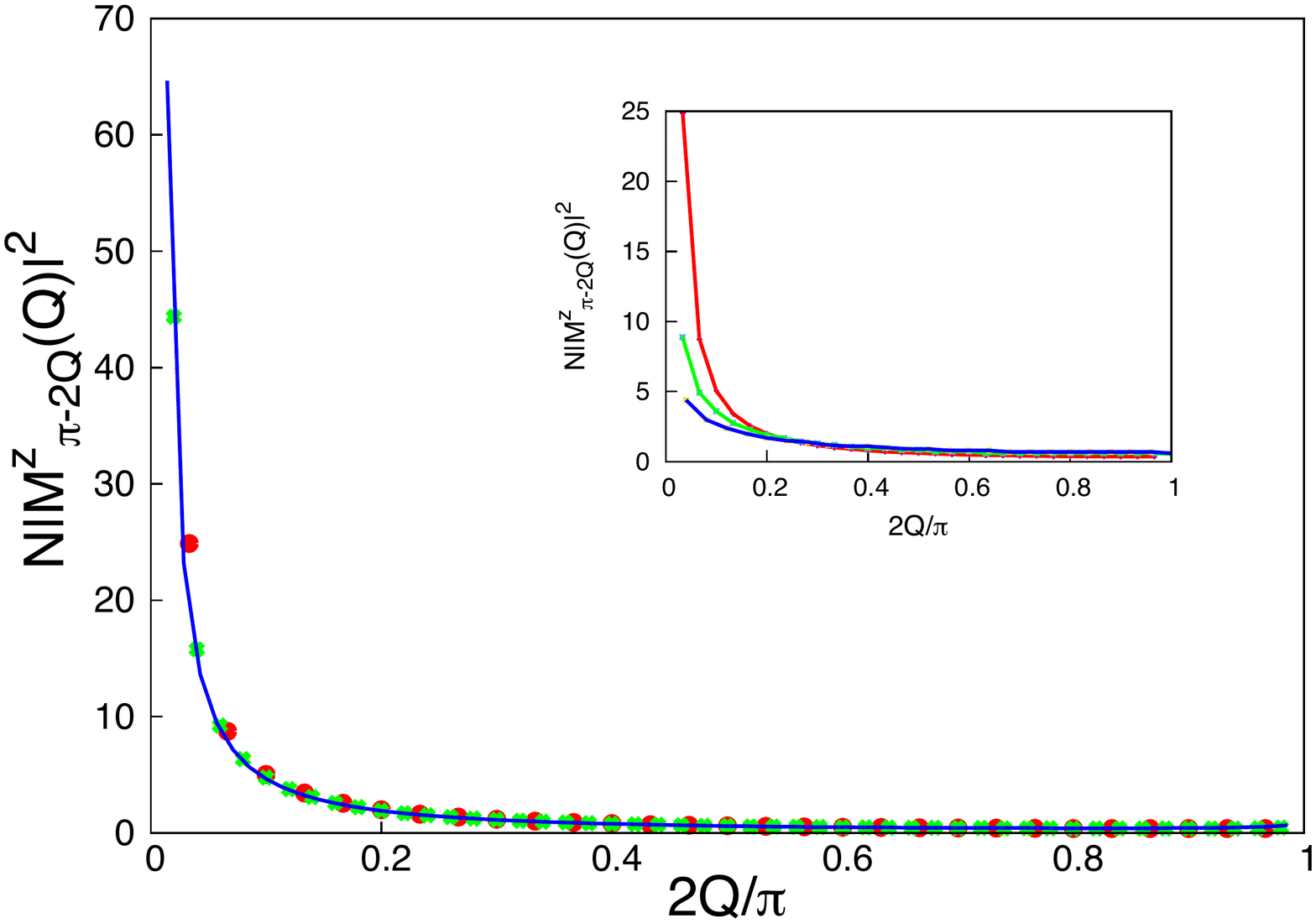}
	\caption{Scaled $N\vert\langle \pi-Q\vert S^z_{\pi-2Q}\vert Q\rangle\vert^2$ 
		versus $Q$ 
		for $\Delta=1$ and various $N$. The matrix elements are symmetric with 
		respect to $Q=\pi/2$. The inset shows $N\vert\langle \pi-Q\vert 
		S^z_{\pi-2Q}\vert Q\rangle\vert^2$ versus $Q$ for $\Delta=1,0.6,0.4$.} 
	\label{fig2s}
\end{figure}

The dominant matrix elements for the longitudinal magnetic potential 
are characterized by a $\pi$-transfer process $\vert \langle Q+
\pi\vert S^z_\pi\vert Q\rangle\vert^2$ and by a velocity-flipping 
$\vert \langle \pi-Q\vert S^z_{\pi-2Q}\vert Q\rangle\vert^2$ 
process in the same branch of the one-spinon spectrum.
We have numerically evaluated the $\pi$-transfer and the same branch 
velocity-flipping  matrix elements as depicted in 
Figs.\ref{fig1s} and \ref{fig2s}, scaling as,
\bea
\vert \langle Q+\pi\vert S^z_\pi\vert Q\rangle\vert^2 &\simeq& {f^z(Q)\over 
	N^{2\mathcal{Z}^2-1}}
\nonumber\\
\vert \langle \pi-Q\vert S^z_{\pi-2Q}\vert Q\rangle\vert^2 &\simeq& {h^z(Q)\over N},
\eea
\noindent
and determined $f^z(Q),h^z(Q)$. Numerically, $f^z(Q)$ is an almost 
constant function with respect to the spinon momentum $Q$ 
(inset Fig. \ref{fig4s}), while $h^z(Q)$ 
is a rapidly decreasing one. Note that the scaling of this $\pi$ 
transfer matrix element has been analytically studied by 
Kitanine et al. in \cite{kitanine1,kitanine2}.

\subsubsection{Transverse potential $V=g S^x_n$}

\begin{figure}[h!]
	\centering
	\includegraphics[width=0.75\linewidth]{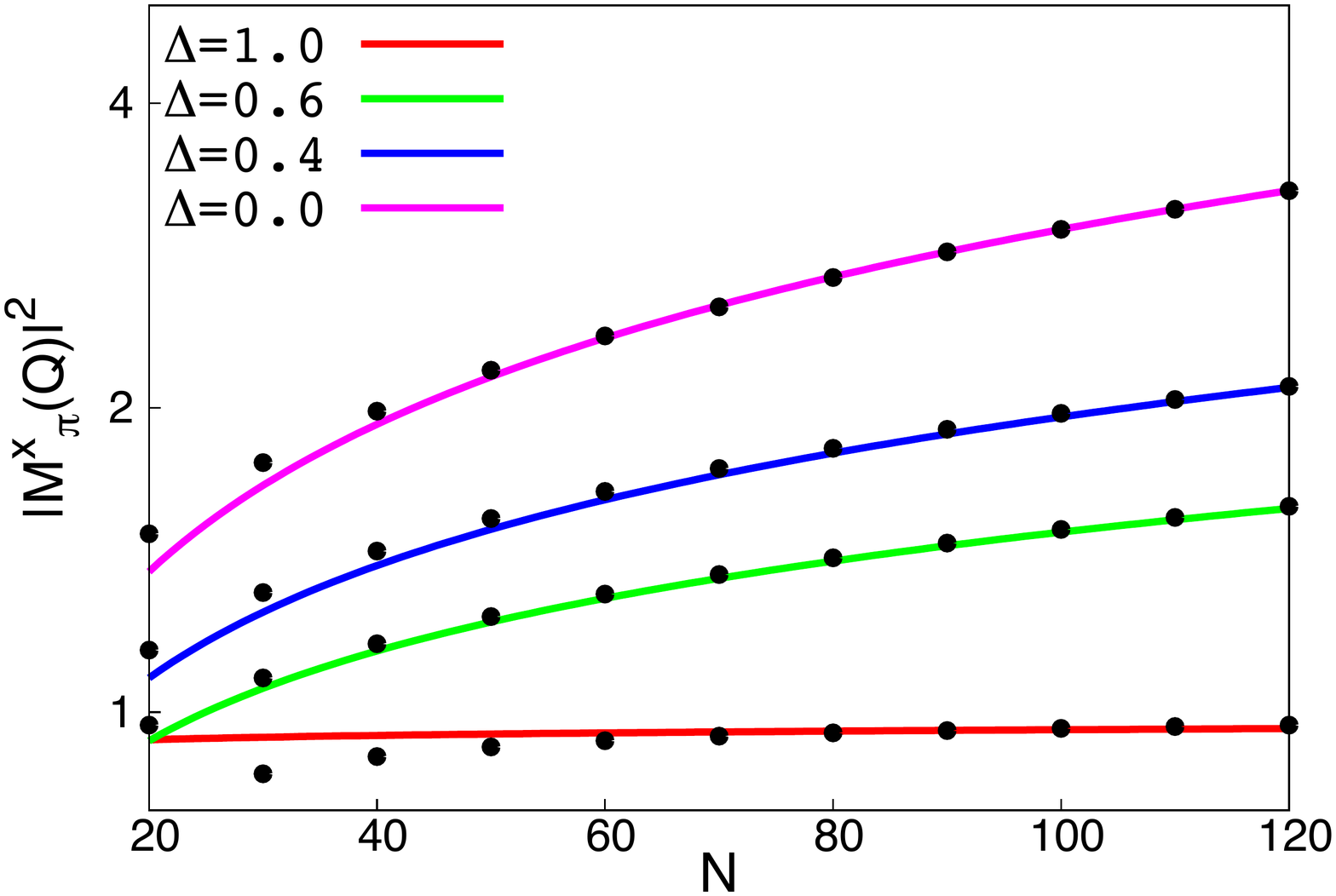}
	\caption{$\vert \langle Q+\pi\vert S^x_\pi\vert Q\rangle\vert^2$ 
		versus $N$ for $\Delta=1,0.6,0.4,0$ for $Q=2\pi/10$. The solid lines 
		are the fitted curve of the form $\vert \langle Q+\pi\vert S^x_\pi\vert 
		Q\rangle\vert^2\simeq {1/N^{{1\over 2\mathcal{Z}^2}-1}}$ 
		while the dots are the numerical data.} 
	\label{fig3s}
\end{figure}

\begin{figure}[h!]
	\centering
	\includegraphics[width=0.75\linewidth]{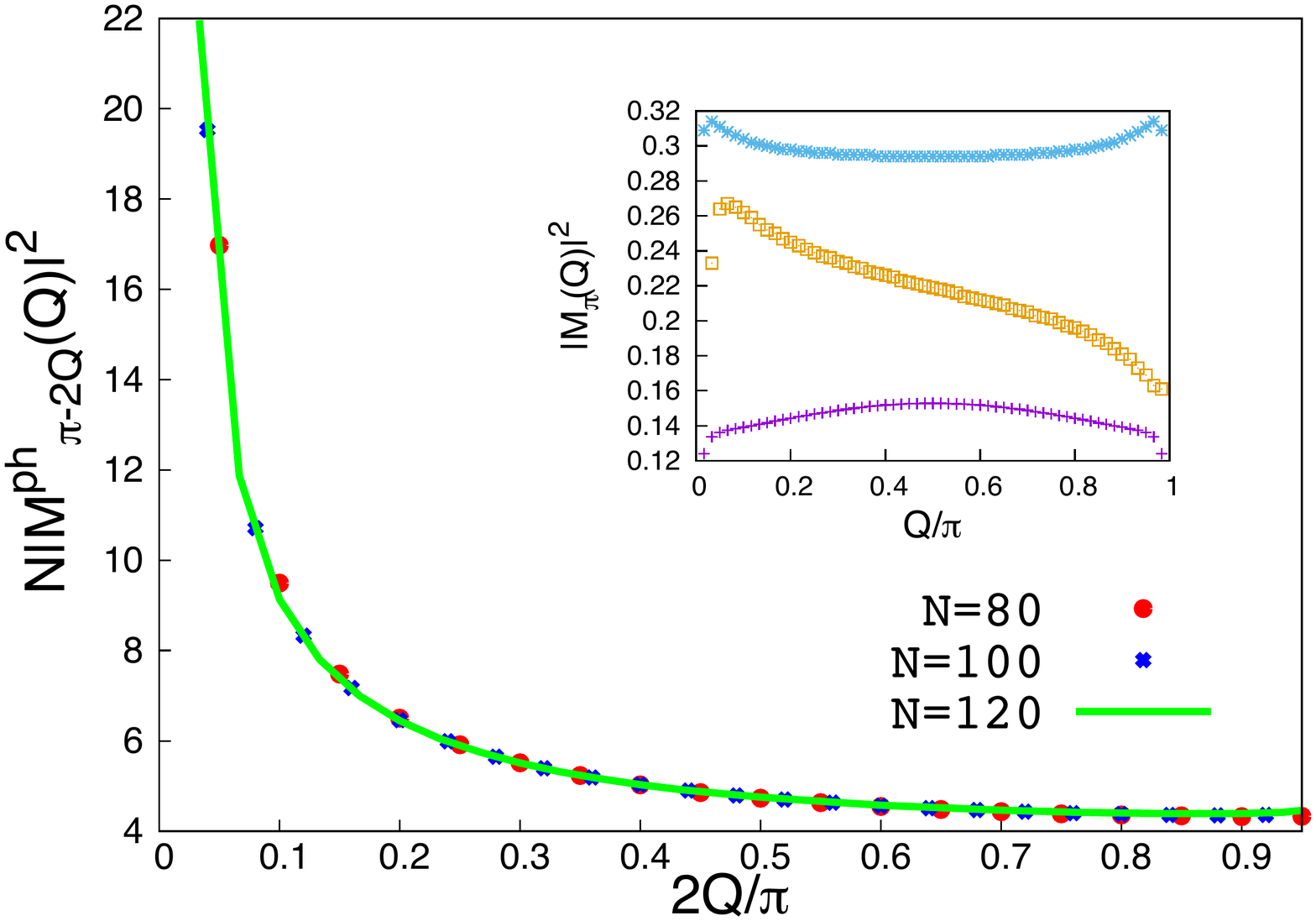}
	\caption{$N\vert \langle\pi-Q\vert V_{\pi-2Q}\vert Q\rangle\vert^2$ 
		versus $Q$ for $\Delta=1$. 
		The inset shows the $\pi$-transition versus $Q$ 
		for the longitudinal $\vert M^z_{\pi}(Q)\vert^2$ (light blue), 
		transverse  $\vert M^x_{\pi}(Q)\vert^2$ (orange) and spin-phonon interaction  
		$\vert M^{ph}_{\pi}(Q)\vert^2$ (purple) for $N=120$ and $\Delta=1$.} 
	\label{fig4s}
\end{figure}

The dominant matrix elements for the transverse magnetic potential 
are characterized by a $\pi$-transfer $\vert \langle Q+\pi\vert 
S^x_\pi\vert Q\rangle\vert^2$ which 
we numerically evaluate as depicted in Fig.\ref{fig3s} 
and scales as \cite{kitanine1,kitanine2}, 
\be 
\vert \langle Q+\pi\vert S^x_\pi\vert Q\rangle\vert^2\simeq {f^x(Q)\over 
	N^{{1\over 2\mathcal{Z}^2}-1}}.
\ee
\noindent
Note for values of $\Delta$ close to one, $f^{x}(Q)$ can be considered 
constant (inset Fig. \ref{fig4s}), while for values of $\Delta$ 
close to zero it is a monotonically decreasing function.

\subsubsection{Spin-phonon interaction}

Finally we discuss a spin-phonon interaction. Similarly with the 
longitudinal magnetic potential, the dominant matrix elements are 
characterized by a $\pi$-transfer $\vert \langle Q+\pi \vert V_\pi\vert 
Q\rangle\vert^2$ and by a velocity-flipping 
$\vert \langle \pi-Q\vert V_{\pi-2Q}\vert Q\rangle\vert^2$ in the same branch 
of the one-spinon spectrum.

Similarly, we obtain that the $\pi$-transfer and the same branch velocity 
flipping  matrix elements behave as,
\bea 
\vert \langle Q+\pi\vert V_\pi\vert Q\rangle\vert^2&\simeq& {f^{ph}(Q)\over 
	N^{\alpha}}
\nonumber\\
\vert \langle \pi-Q\vert V_{\pi-2Q}\vert Q\rangle\vert^2
&\simeq& {h^{ph}(Q)\over N},
\eea
with $\alpha\simeq 0.4$ for $Q=2\pi/10$ and small corrections 
with respect to $Q$. Note that similarly to the longitudinal potential, 
$h^{ph}(Q)$ is a rapidly decreasing function for $\Delta>0$ while 
for $\Delta=0$ is constant and equal to one. Moreover, $f^{ph}(Q)$ and 
$h^{ph}(Q)$ are symmetric with respect to $\pi/2$.

\subsection{Scattering Theory }

In this section we give a short description of the T-matrix approach 
for evaluating  the transmission and reflection probability of spinon 
scattering from magnetic and non-magnetic potentials.
The Lippmann-Schwinger equation is 
\be 
\vert \psi\rangle=\vert \psi_0\rangle+(E-H_0)^{-1}V\vert \psi\rangle,
\label{eq5}
\ee
where $\vert \psi_0\rangle$ is the unperturbed incident eigenstate of 
the $H_0$ Hamiltonian and $V$ is the scattering potential. 
By iteratively solving the Lippmann-Schwinger equation we formally obtain 
the Born Series
\be  
\vert {\psi}\rangle=\sum_{k=0}^\infty (G_0V)^k\vert\psi_0\rangle.
\ee
\noindent
$G_0$ is the Green's function defined as
\be 
G_0(E)=\lim_{\epsilon\rightarrow 0} 
{1\over E-H_0+i\epsilon}=\lim_{\epsilon\rightarrow 0}
\sum_{\lbrace\lambda\rbrace} {\vert \lbrace\lambda\rbrace\rangle 
	\langle\lbrace\lambda\rbrace\vert \over E-E_{\lbrace\lambda\rbrace}+i\epsilon}
\label{eq6}
\ee

\noindent
and $\vert\lbrace \lambda\rbrace\rangle$ denotes the Hamiltonian $H_0$ 
eigenstate parametrized by the set of parameters $\lbrace\lambda\rbrace$.

Next, we define the T-matrix, a "black box" that contains all the 
relevant information about the scattering process
\be 
T=V(1-G_0V)^{-1}=V\sum_{k=0}^\infty (G_0V)^k,
\label{eq7}
\ee

\noindent
from which we obtain the "diagonal" transmission probability (probability 
to find the particle in the same state) of a 
particle in the state $\vert Q \rangle$ with energy $\varepsilon_{Q}$ 
and group velocity $u_Q={d\varepsilon_{Q}/d Q}$ \cite{merz}
\be 
\mathcal{T}_{Q,Q}=(1+{N\over u_Q}\Im\langle Q \vert 
T\vert Q\rangle)^2+
({N\over u_Q}\Re\langle Q \vert T\vert Q \rangle)^2.
\label{eq8}
\ee
\noindent

If we apply this formalism to the case of the anisotropic Heisenberg 
chain we obtain the Green's function for the single spinon, 
\be 
G_0(Q)=\lim_{\epsilon\rightarrow 0}\sum_{\lbrace\lambda\rbrace} 
{\vert \lbrace\lambda\rbrace\rangle \langle\lbrace\lambda\rbrace\vert 
	\over \varepsilon_Q-E_{\lbrace\lambda\rbrace}+i\epsilon}, 
~~\varepsilon_Q=\frac{\pi}{2}\frac{\sin\gamma}{\gamma}|\sin Q|.
\label{eq9}
\ee  

\noindent
Moreover, in the case of an even spin chain, since it is impossible to 
physically distinguish between two spinon excitations that have the same 
energy and group velocity but live in different branches of the 
Cloiseaux-Pearson spectrum, we also define the "non-diagonal" transmission 
probability as
\be 
\mathcal{T}_{Q,Q+\pi}=({N\over u_Q}\Im\langle Q+\pi \vert 
T\vert Q\rangle)^2+
({N\over u_Q}\Re\langle Q+\pi \vert T\vert Q \rangle)^2.
\label{eq10}
\ee
\noindent
In the odd site case, at least in the lower branch approximation, 
this is not needed.

Let us analytically evaluate the transmission probability for a "free" 
spinon (including only one-spinon excitations) toy model 
in an odd site spin chain 
with a potential matrix
\be 
V={g\over N^{\alpha}} \sum_{Q,Q^{'}} \vert Q\rangle \langle Q'\vert, 
\label{eq11}
\ee
where $\alpha>0$ and $\vert Q\rangle,\vert Q'\rangle$ are one-spinon states. 
Note that $\alpha=1$ 
corresponds to the case of a $\delta$-like potential which we presented 
in the main text.
Using eqs.(\ref{eq7}-\ref{eq10}) we evaluate the $T$ matrix 
and the transmission amplitude $\mathcal{T}$, a function of $g_{eff}/u_Q$, 
$g_{eff}\equiv gN^{1-\alpha}$
\bea 
T(Q,Q')&=& {g\over N^{\alpha}} {1\over {1-(I_1+iI_2)}},
\nonumber\\
I_1&=& {g_{eff}\over 2\pi}P \int {dq\over {\varepsilon_{Q}-\varepsilon_q}}
=-{g_{eff}\over 2\pi \vert u_{Q}\vert} \log ( {1+\vert \cos Q\vert 
	\over 1-\vert \cos Q \vert} ),~~I_2=-{g_{eff}\over \vert u_{Q}\vert},
\nonumber\\
\mathcal{T}_{Q,Q}&=&t_1^2+(1+t_2)^2,
\\
t_1&=&\Big({g_{eff}\over u_Q}\Big){(1-I_1)\over {(1-I_1)^2+(I_2)^2}},~~~
t_2=\Big({g_{eff}\over u_Q}\Big){I_2\over {(1-I_1)^2+(I_2)^2}}.
\nonumber
\label{free_t}
\end{eqnarray}
\noindent

This result holds for $I^2_1+I^2_2<1$. 
An interesting observation is that if we include only the on-shell 
matrix elements, 
i.e. only the $I_2$ part, then we get the correct qualitative behavior 
in the dependence on $N$ and the spinon energy $\varepsilon_{Q}$, 
with the rest of the matrix elements given by $I_1$ acting as corrections 
to the amplitude. Moreover, by repeating the same calculation  
for an even site spin chain, we can technically understand our results 
of the spinon transfer between the two branches, since for $\alpha<1$ 
the transmission probability $\mathcal{T}_{Q,Q}\rightarrow 0$ as $N$ 
increases while in the case $\alpha>1$ the transmission probability 
$\mathcal{T}_{Q,Q}\rightarrow 1$ as $N$ increases and $\mathcal{T}_{Q,Q}$ 
is scale invariant when $\alpha=1$.

Furthermore, using only the dominant matrix elements 
we can obtain a qualitative 
expression for the transmission amplitude of the potentials that we have 
previously discussed. Also we use the symmetry with respect to $Q=\pm \pi/2$, 
which holds for the longitudinal and the spin-phonon interaction, while 
for the transverse potential we notice that although this condition is 
not fulfilled, the same procedure gives essentially the correct result. 
Let us first apply the above for an even site spin chain. 

The diagonal element of the T-matrix is given by

\bea 
T(Q,Q)&=&i{u_{Q}\over N}  \sum_{n=1}^\infty (-1)^n 
\left(1\over 2u_{Q}\right)^{2n}
(\mathcal{F}^{n}+\mathcal{G}^{n}),
\\
\mathcal{F}&\equiv &N(\vert\langle  Q+\pi\vert V_\pi\vert  Q\rangle \vert + 
\vert \langle\pi-Q\vert V_{\pi-2Q}\vert  Q\rangle \vert)^2
\nonumber\\
\mathcal{G}& \equiv &N(\vert\langle  Q+\pi\vert V_\pi\vert Q\rangle \vert - 
\vert \langle \pi-Q\vert V_{\pi-2Q}\vert Q\rangle \vert)^2,  
\nonumber
\label{eq14}
\eea 
\noindent
where we have used that only the even terms contribute to the 
Born series and that 
\be 
\sum_{m=even}^{2n} {2n \choose m} \vert \langle Q+\pi\vert V_\pi
\vert Q\rangle\vert ^{2n-m}\vert \langle \pi-Q \vert V_{\pi-2Q}
\vert Q\rangle\vert^{m} ={1\over 2 N^n} (\mathcal{F}^{n}+\mathcal{G}^{n}).
\label{eq15}
\ee

\noindent
Summing up the series to obtain the diagonal T-matrix element, 
\be 
T(Q,Q)=-i{u_{Q}\over N} \left[{\mathcal{F} \over 4u_{Q}^2+\mathcal{F}}+
{\mathcal{G}\over 4u_{Q}^2+\mathcal{G}}\right]
\label{eq16}
\ee

\noindent
and the transmission amplitude,
\bea 
&\mathcal T_{Q,Q}=\left({16u^4_{Q}-\mathcal{F}\mathcal{G}\over 16u^4_{Q}+
\mathcal{F}\mathcal{G}+4u_{Q}^2(\mathcal{F}+\mathcal{G})}\right)^2.
\label{eq17}
\eea

Additionally we calculate the probability that the spinon is transmitted 
through the second branch.  Using a similar procedure 
we obtain the $T(Q,Q+\pi)$ T-matrix element
\bea 
T(Q,Q+\pi)&=2u^2_Q\left[ {\vert\langle Q+\pi\vert V_\pi\vert Q\rangle\vert
+ \vert\langle \pi-Q\vert V_{\pi-2Q}\vert Q\rangle\vert\over {4 u^2_Q
	+\mathcal{F}}}\right.\\ \nn
&+\left.{\vert\langle Q+\pi\vert V_\pi\vert Q\rangle\vert-\vert\langle 
\pi-Q\vert V_{\pi-2Q}\vert Q\rangle\vert\over {4 u^2_Q+\mathcal{G}}}\right],
\eea
and $\mathcal{T}_{Q,Q+\pi}=({N\over u_Q})^2 T(Q,Q+\pi)^2$.
\\
\\
In the case of an odd site spin chain we have a much simpler situation, 
since the $\pi$-transfer matrix element and the second branch are non-existent.

Finally note that, when $\varepsilon_{Q}$ is not close to zero, the quantity 
$N\langle \pi-Q\vert S^z_{\pi-2Q}\vert Q\rangle=f^z(Q)$ can be 
considered negligible. Therefore by taking the logarithm of the transmission 
coefficient and using the identity $\log(1-x)= x-x^2/2+O(x^3)$, we obtain 
that for 
$(g_{eff}/u_Q)^2/4f^z(Q)<<1$,
\be 
\log\mathcal{T}_{Q,Q}\simeq -f^z(Q) ({g_{eff}\over u_Q})^2\Rightarrow 
\mathcal T_{Q,Q}\simeq e^{-f^z(Q) ({g_{eff}\over u_Q})^2},~~~ g_{eff}\equiv 
gN^{1-\mathcal{Z}^2}.
\label{eq18}
\ee

\end{document}